\documentclass[pre,superscriptaddress,showpacs]{revtex4}
\usepackage{amsmath,amsfonts,graphicx,psfrag}

\begin{document}
\title{Hydrodynamics of probabilistic ballistic annihilation}
\author{Fran\c{c}ois Coppex}
\affiliation{Department of Theoretical Physics, University of Gen\`eve, 
CH-1211 Gen\`eve 4, Switzerland}
\author{Michel Droz}
\affiliation{Department of Physics, University of Gen\`eve, 
CH-1211 Gen\`eve 4, Switzerland}
\author{Emmanuel Trizac}
\affiliation{Laboratoire de Physique Th\'eorique (UMR 8627 du CNRS), B\^atiment 210, 
Universit\'e de Paris-Sud, 91405 Orsay, France}
\begin{abstract}
We consider a dilute gas of hard spheres in dimension $d \geq 2$ that upon
collision either annihilate with probability $p$ or undergo an elastic
scattering with probability $1-p$. For such a system neither mass, momentum,
nor kinetic energy are conserved quantities. We establish the hydrodynamic
equations from the Boltzmann equation description. Within the Chapman-Enskog
scheme, we determine the transport coefficients up to Navier-Stokes order, and
give the closed set of equations for the hydrodynamic fields chosen for the
above coarse grained description (density, momentum and kinetic
temperature). Linear stability analysis is performed, and the conditions of
stability for the local fields are discussed.
\end{abstract}
\maketitle

%=========================================================
\section{Introduction}\label{section1}
%=========================================================

The hydrodynamic description of a low density gas of elastic hard spheres supported by an underlying kinetic theory attracted a lot of attention already more than 40 years ago~\cite{bogolubov,cohen,chapmancowling}. It has now become a well established description~\cite{santosmontaneroduftybrey}. 
A key ingredient in the hydrodynamic approach is the existence of collisional invariants (quantities conserved by the --instantaneous--
collisions). The question of the relevance of a coarse-grained hydrodynamic 
approach is therefore more problematic when the kinetic energy is no longer a 
collisional invariant \cite{goldhirsch}. This is the case of rapid granular 
flows (that may be modeled by inelastic hard spheres in a first approach), 
where the hydrodynamic picture, despite including a hydrodynamic field 
associated with the kinetic energy density, is nevertheless reliable 
(see, e.g., \cite{BreyCub,GarMon,Meerson,Matthieu} and Dufty for a 
review \cite{duftycondmat}). It seems natural to test hydrodynamic-like 
approaches further and in more extreme conditions, and investigate a 
system where particles react so that there exist {\em no collisional 
invariants}. The present article, focusing on the derivation of
the hydrodynamic description for such a system, 
is a first step in this direction. 

The system we consider is made of an assembly of hard spheres that move 
ballistically between collisions. Whenever two particles meet, they either 
annihilate with probability $p$, or undergo an elastic collision with 
probability $(1-p)$. The model of probabilistic ballistic annihilation in 
one dimension for bimodal discrete initial velocity distributions was 
introduced in~\cite{richardsonkafri}, whereas for higher dimensions and 
arbitrary continuous initial velocity distributions it was considered in~\cite{coppex}. When $p=1$, we recover the annihilation model originally defined by Elskens and Frisch~\cite{elskensfrisch}, that has attracted some
attention since 
\cite{BenNaim,piasecki,reydrozpiasecki,Blythe,Krapivsky,Lettre,trizac,physica}.
In one dimension (again for $p=1$), the problem is well understood for
discrete initial velocity distributions~\cite{piasecki,reydrozpiasecki}. On the contrary, higher dimensions introduce complications that make the problem much more difficult to treat~\cite{Lettre,trizac}. Only a few specific initial velocity conditions lead to systems that are tractable using the standard tools of kinetic theory~\cite{physica}. 

Our starting point will be the Boltzmann equation, which describes correctly
the low density limit of granular gases (see \cite{breynew} and 
\cite{resibois,lanford} for the elastic case). For annihilation
dynamics, the ratio of particle diameter to mean free path vanishes in the
long-time limit, such that for $d>1$ the Boltzmann equation is valid at least
at late times~\cite{trizac,physica}. 
For such a dynamics, none of the standard hydrodynamic fields (i.e., mass,
momentum, and energy) is associated with a conserved quantity. There are
therefore three nonzero decay rates, one for each field. Numerical evidences
show that in the long-time limit, a non-Maxwellian scaling solution for the homogeneous
system appears (homogeneous cooling state (HCS)~\cite{Lettre,trizac}, which also
exists for inelastic hard spheres
\cite{breymonterocubero2}). Nothing is known about the stability of the latter
solution,
the only existing result being that in one dimension, with a bimodal initial
velocity distribution, clusters of particles are spontaneously formed by the
dynamics~\cite{reydrozpiasecki}. In view of this situation we develop a hydrodynamic description for probabilistic ballistic annihilation. The limiting case of vanishing annihilation probability $p \to 0$ gives the known results for elastic dilute gases~\cite{premier}, whereas the other limit $p \to 1$ yields the case of pure annihilation.

In order to derive the hydrodynamic equations, we make use of the Chapman-Enskog method. We thus consider (at least) two distinct timescales in the system. The microscopic timescale is characterized by the average collision time and the corresponding length scale defined by the mean free path. The macroscopic timescale is characterized by the typical time of evolution of the hydrodynamic fields and their inhomogeneities. The fact that those two timescales are very different implies that on the microscopic timescale the hydrodynamic fields vary only slightly. Therefore they are on such length- and timescales only very weakly inhomogeneous. Combined with the existence of a normal solution for the velocity distribution function (i.e., a solution such that all time dependence may be expressed through the hydrodynamic fields), 
this allows for a series expansion in orders of the gradients, i.e., application of the Chapman-Enskog method. 
The knowledge of the hydrodynamic equations thus obtained to first order 
allows us to perform a stability analysis. Taking the HCS as a reference state, 
we study the corresponding small spatial deviations of the hydrodynamic fields. 
%This allows for a linearization of the hydrodynamic equations, 
%and to perform a linear stability analysis. 

The paper is organized as follows. In Sec.~\ref{section2} we present the
Boltzmann equation for probabilistic ballistic annihilation, and establish the
subsequent balance equations. Sec.~\ref{section3} is devoted to the
Chapman-Enskog solution of the balance equations. For this purpose we consider
an expansion of the latter equations in a small formal parameter. The solution
to zeroth order provides the hydrodynamic fields of the HCS. Assuming small
spatial inhomogeneities, we make use of an explicit normal solution for the
velocity distribution function to first order. This allows us to establish the
expression for the transport coefficients and for the decay rates to first
order, and thus the closed set of equations for the hydrodynamic fields. 
The technical aspects of the derivations are presented in the Appendixes
while our main results are gathered in Eqs. (\ref{balanceequationsend}). In
Sec.~\ref{section4}, we study the linear stability of those equations
around the HCS. Finally, Sec.~\ref{section5} contains the discussion of the
results and our conclusions. Since from the point of view of dissipation
probabilistic ballistic annihilation shares some features with granular gases,
making several parallels between those two systems will prove to be
instructive. 

%=========================================================
\section{Boltzmann and balance equations}\label{section2}
%=========================================================

The Boltzmann equation for the one particle distribution function $f(\mathbf{r},\mathbf{v};t)$ of a low-density system of hard spheres annihilating with probability $p$ is given by
\begin{equation}
(\partial_t + \mathbf{v}_1 \cdot \boldsymbol{\nabla}) f(\mathbf{r},\mathbf{v}_1;t) = p J_a[f,f] + (1-p) J_c[f,f], \label{boltzmann}
\end{equation}
where the annihilation operator $J_a$ is defined by~\cite{trizac}
\begin{equation}
J_a[f,g] = - \sigma^{d-1} \beta_1 \int_{\mathbb{R}^d} d \mathbf{v}_2 \, v_{12} f(\mathbf{r},\mathbf{v}_2;t) g(\mathbf{r},\mathbf{v}_1;t)\label{definitionannihilation}
\end{equation}
and the elastic collision operator $J_c$ is defined by~\cite{grad,lanford,breynew}
\begin{equation}
J_c[f,g] = \sigma^{d-1} \int_{\mathbb{R}^d} d \mathbf{v}_2 \int d \widehat{\boldsymbol{\sigma}} \,( \widehat{\boldsymbol{\sigma}} \cdot \mathbf{v}_{12}) \theta(-  \widehat{\boldsymbol{\sigma}} \cdot \mathbf{v}_{12}) (b^{-1} -1) g(\mathbf{r},\mathbf{v}_1;t) f(\mathbf{r},\mathbf{v}_2;t).
\end{equation}
In the above expressions, $\sigma$ is the diameter of the particles, $\mathbf{v}_{12} = \mathbf{v}_1 - \mathbf{v}_2$ the relative velocity, $v_{12} = |\mathbf{v}_{12}|$, $\theta$ is the Heaviside distribution, $\widehat{\boldsymbol{\sigma}}$ a unit vector joining the centers of two particles at collision and the corresponding integral is running over the solid angle, $\beta_1 = \pi^{(d-1)/2} / \Gamma[(d+1)/2]$ where $\Gamma$ is the gamma function, and $b^{-1}$ an operator acting on the velocities as follows~\cite{ggnoije}:
\begin{equation}
b^{-1} \mathbf{v}_{12} = \mathbf{v}_{12} - 2 (\mathbf{v}_{12} \cdot \widehat{\boldsymbol{\sigma}}) \widehat{\boldsymbol{\sigma}},
\end{equation}
\begin{equation}
b^{-1} \mathbf{v}_{1} = \mathbf{v}_{1} - (\mathbf{v}_{12} \cdot \widehat{\boldsymbol{\sigma}}) \widehat{\boldsymbol{\sigma}}.
\end{equation}
Since $J_c$ describes elastic collisions, this operator conserves the mass, momentum, and energy. On the other hand, $J_a$ describes the annihilation process and thus none of the previous quantities are conserved. 

In order to write hydrodynamic equations, we need to define the following local hydrodynamic fields:
\begin{subequations}
\begin{eqnarray}
n(\mathbf{r},t) &=& \int_{\mathbb{R}^d} d \mathbf{v} \, f(\mathbf{r},\mathbf{v};t), \label{defn} \\
\mathbf{u}(\mathbf{r},t) &=& \frac{1}{n(\mathbf{r},t)} \int_{\mathbb{R}^d} d \mathbf{v} \, \mathbf{v} f(\mathbf{r},\mathbf{v};t),\label{defu} \\
T(\mathbf{r},t) &=& \frac{m}{n(\mathbf{r},t) k_B d} \int_{\mathbb{R}^d} d \mathbf{v} \, \mathbf{V}^2 f(\mathbf{r},\mathbf{v};t). \label{deft}
\end{eqnarray}
\end{subequations}
where $n(\mathbf{r},t)$, $\mathbf{u}(\mathbf{r},t)$, and $T(\mathbf{r},t)$ are the local number density, velocity, and temperature, respectively. The definition of the temperature follows from the principle of equipartition of energy. In Eq.~(\ref{deft}), $k_B$ is the Boltzmann constant and $\mathbf{V} = \mathbf{v} - \mathbf{u}(\mathbf{r},t)$ is the deviation from the mean flow velocity. The balance equations follow from integrating the moments $1$, $m \mathbf{v}$, and $m v^2/2$ with weight given by the Boltzmann equation~(\ref{boltzmann}). We thus obtain
\begin{subequations}
\begin{eqnarray}
& & \partial_t n + \nabla_i (n u_i) = - p \omega[f,f], \\
& & \partial_t u_i + \frac{1}{m n} \nabla_j P_{ij} + u_j \nabla_j u_i = - p \frac{1}{n} \omega[f,V_i f], \qquad i=1,\ldots,d, \\
& & \partial_t T + u_j \nabla_j T + \frac{2}{n k_B d}( P_{ij} \nabla_i u_j + \nabla_j q_j) = p \frac{T}{n} \omega[f,f] - p \frac{m}{n k_B d} \omega[f,V^2 f],
\end{eqnarray}\label{balanceequations}
\end{subequations}
where we have summation over repeated indices, $\mathbf{u}=(u_1, \ldots , u_d)$, and
\begin{equation}
\omega[f,g] = \sigma^{d-1} \beta_1 \int_{\mathbb{R}^{2d}} d \mathbf{v}_1 d \mathbf{v}_2 |\mathbf{v}_{12}| g(\mathbf{r}_1,\mathbf{v}_1;t) f(\mathbf{r}_2,\mathbf{v}_2;t). \label{definitionomega}
\end{equation}
In the balance equations~(\ref{balanceequations}), the pressure tensor $P_{ij}$ and heat-flux $q_i$ are defined by
\begin{equation}
P_{ij}(\mathbf{r},t) = m \int_{\mathbb{R}^d} d \mathbf{v} \, V_i V_j f(\mathbf{r},\mathbf{v};t) =  \int_{\mathbb{R}^d} d \mathbf{v} \, f(\mathbf{r},\mathbf{v};t) D_{ij}(\mathbf{V}) + \frac{n}{\beta} \delta_{ij}, \label{pressure}
\end{equation}
\begin{equation}
q_i(\mathbf{r},t) = \int_{\mathbb{R}^d} d \mathbf{v} S_i(\mathbf{V}) f(\mathbf{r},\mathbf{v};t), \label{heat}
\end{equation}
where $\beta=1/(k_B T)$ and 
\begin{equation}
D_{ij}(\mathbf{V}) = m \left( V_i V_j - \frac{V^2}{d} \delta_{ij} \right), \label{definitiondij}
\end{equation}
\begin{equation}
S_i(\mathbf{V}) = \left( \frac{m}{2} V^2 - \frac{d+2}{2} k_B T \right) V_i.\label{definitionsi}
\end{equation}
One sees from Eqs.~(\ref{balanceequations}) that when the annihilation
probability $p \to 0$, all quantities are conserved. In addition, the long
time solution of the system in this limit is given by the Maxwell distribution~\cite{coppex}.

%=========================================================
\section{Chapman-Enskog solution}\label{section3}
%=========================================================
In order to solve Eqs.~(\ref{balanceequations}), it is necessary to obtain a closed set of equations for the hydrodynamic fields. This can be done using the Chapman-Enskog method, by expressing the functional dependence of the pressure tensor $P_{ij}$ and of the heat flux $q_i$ in terms of the hydrodynamic fields. Note that other routes have been developed as well~\cite{ramirez,grad}. A thorough comparison of the different approaches seems, however, to be a difficult attempt~\cite{ramirez}. In order to apply the Chapman-Enskog method, it is necessary to make two assumptions. The first one is that all temporal and spatial dependence of the distribution function $f(\mathbf{r},\mathbf{v};t)$ may be expressed as a functional dependence on the hydrodynamic fields:
\begin{equation}
f(\mathbf{r},\mathbf{v};t) = f\left[ \mathbf{v},n(\mathbf{r},t),\mathbf{u}(\mathbf{r},t),T(\mathbf{r},t) \right].\label{normalsolution}
\end{equation}
What is the physical justification for the existence of such a \emph{normal
  solution}? Suppose that the variations of the hydrodynamic fields are small
on the scale of the mean free path, for example $(n \sigma^{d-1})^{-1}
|\boldsymbol{\nabla} \ln n | \ll 1$. Therefore, to first order the functional
dependence of the distribution function may be made local in the hydrodynamic
fields, leading to the normal solution written above. Note that none of the
hydrodynamic fields is associated with a conserved quantity. The theoretical
question that arises is to know if the new timescales thereby introduced by
the cooling rates are shorter than what is allowed for the existence of a
normal solution~\cite{duftycondmat}. For sufficiently small $p$ this should be
the case. However, in the related context of granular gases,
this point is not yet quantitatively clarified and is still subject to discussions~\cite{duftycondmat,tangoldhirsch,breyruizmonterocubero}. The justification of the normal solution may be done \emph{a posteriori} by studying the relevance of the results through the appearance of the HCS, for example~\cite{trizac,breymonterocubero2}. The second assumption is based on the existence of (at least) two distinct timescales. The microscopic timescale is characterized by the average collision time and the spatial length which is defined by the corresponding mean free path. On the other hand, the macroscopic timescale is defined by a typical timescale describing the evolution of the hydrodynamic fields and their inhomogeneities. The difference in those two timescales implies that on the microscopic timescale the hydrodynamic fields vary only very slightly. Thus, those fields are on such time and space scales only very weakly inhomogeneous. 
This allows for a series expansion in orders of the gradients of the fields:
\begin{equation}
f = f^{(0)} + \lambda f^{(1)} + \lambda^2 f^{(2)} + \ldots,\label{spatialhierarchy}
\end{equation}
where each power of the formal small parameter $\lambda$ means a given order in a spatial gradient. The formal parameter $\lambda$ may be seen as the ratio of the mean free path to the wavelength of the variation of the hydrodynamic fields. This shows again the idea of the separation of both microscopic and macroscopic time and length scales. The Chapman-Enskog method assumes the existence of a timescale hierarchy, and thus of a time derivative hierarchy as well:
\begin{equation}
\frac{\partial}{\partial t} = \frac{\partial^{(0)}}{\partial t} + \lambda \frac{\partial^{(1)}}{\partial t} + \lambda^2 \frac{\partial^{(2)}}{\partial t} + \ldots,\label{temporalhierarchy}
\end{equation}
where a given order in the temporal hierarchy~(\ref{temporalhierarchy}) corresponds to the same order in the spatial hierarchy~(\ref{spatialhierarchy}). One thus concludes that the higher the order of the spatial gradient, the slower the corresponding temporal variation. Inserting expansions~(\ref{spatialhierarchy}) and~(\ref{temporalhierarchy}) in the Boltzmann equation~(\ref{boltzmann}) one obtains
\begin{equation}
\left( \sum_{k \geq 0} \lambda^k \frac{\partial^{(k)}}{\partial t} + \mathbf{v}_1 \cdot \boldsymbol{\nabla} \right) \sum_{l \geq 0} \lambda^{l} f^{(l)} = p J_a\left[\sum_{l \geq 0} \lambda^{l} f^{(l)},\sum_{l \geq 0} \lambda^{l} f^{(l)} \right] + (1-p) J_c \left[ \sum_{l \geq 0} \lambda^{l} f^{(l)}, \sum_{l \geq 0} \lambda^{l} f^{(l)}\right]. \label{expansion}
\end{equation}
Collecting the terms of a given order in $\lambda$ and solving the equations order by order allows us to build the Chapman-Enskog solution.

\subsection{Zeroth order}
%-----------------------------------------
To zeroth order in the gradients, Eq.~(\ref{expansion}) gives
\begin{equation}
\partial_t^{(0)} f^{(0)} = p J_a [ f^{(0)},f^{(0)} ] + (1-p) J_c [ f^{(0)},f^{(0)} ]. \label{eqorder0}
\end{equation}
This equation has a solution, describing the HCS, and which obeys the scaling relation
\begin{equation}
f^{(0)}(\mathbf{r},\mathbf{v};t) = \frac{n(t)}{v_T(t)^d} \widetilde{f}(c). \label{eqscaling}
\end{equation}
The approximate expression for $\widetilde{f}(c)$ was established 
in~\cite{coppex} and is recalled by Eq.~(\ref{f0sonine}). 
In Eq.~(\ref{eqscaling}), $v_T = [2/(\beta m)]^{1/2}$ is the time dependent
thermal velocity, and $c=V/v_T$, $\mathbf{V} = \mathbf{v} - \mathbf{u}$. The existence of a scaling solution of the form~(\ref{eqscaling}) seems to be a general feature that is confirmed numerically (direct Monte-Carlo simulations or molecular dynamics) not only for ballistic annihilation~\cite{trizac} or granular gases~\cite{santosrefereeemmanuel}, but for the dynamics of ballistic aggregation as well~\cite{aggregation,Trizac_coal}.

The function $f^{(0)}$ is isotropic. Thus to this order the pressure tensor~(\ref{pressure}) becomes $P_{ij}^{(0)} = p^{(0)} \delta_{ij}$, where $p^{(0)} = n k_B T$ is the hydrostatic pressure, and the heat flux~(\ref{heat}) becomes $\mathbf{q}^{(0)} = 0$. 

The balance equations~(\ref{balanceequations}) to zeroth order read
\begin{subequations}
\begin{eqnarray}
\partial_t n &=& - p n \xi_n^{(0)}, \label{balanceequations0a} \\
\partial_t u_i &=& - p v_T \xi_{u_i}^{(0)}, \qquad i=1,\ldots,d, \label{balanceequations0b} \\
\partial_t T &=& - p T \xi_T^{(0)}, \label{balanceequations0c}
\end{eqnarray}\label{balanceequations0}
\end{subequations}
where the decay rates are
\begin{subequations}
\begin{eqnarray}
\xi_n^{(0)} &=& \frac{1}{n} \omega[f^{(0)},f^{(0)}], \label{decay0n} \\
\xi_{u_i}^{(0)} &=& \frac{1}{n v_T} \omega[f^{(0)},V_i f^{(0)}], \qquad i=1,\ldots,d, \label{decay0u} \\
\xi_T^{(0)}&=& \frac{m}{n k_B T d} \omega[f^{(0)}, V^2 f^{(0)}] - \frac{1}{n} \omega[f^{(0)},f^{(0)}],\label{decay0t}
\end{eqnarray}\label{decay0}
\end{subequations}
For antisymmetry reasons, one sees from Eq.~(\ref{decay0u}) that $\xi_{u_i}^{(0)}=0$. The two other decay rates are given later on by Eqs.~(\ref{decay0end}).

\subsection{First order}
%-----------------------------------------
To first order in the gradients, the Boltzmann equation~(\ref{expansion}) reads
\begin{equation}
[ \partial_t^{(0)} + J ] f^{(1)} = - [ \partial_t^{(1)} + \mathbf{v}_1 \cdot \boldsymbol{\nabla} ] f^{(0)}, \label{boltzmann1}
\end{equation}
where
\begin{equation}
J f^{(1)} = p L_a[f^{(0)},f^{(1)}] + (1-p) L_c[f^{(0)},f^{(1)}],\label{opcollision}
\end{equation}
with
\begin{equation}
L_a[f^{(0)},f^{(1)}] = -J_a[f^{(0)},f^{(1)}] - J_a[f^{(1)},f^{(0)}],\label{definitionla}
\end{equation}
\begin{equation}
L_c[f^{(0)},f^{(1)}] = -J_c[f^{(0)},f^{(1)}] - J_c[f^{(1)},f^{(0)}].
\end{equation}

The balance equations~(\ref{balanceequations}) to first order become
\begin{subequations}
\begin{eqnarray}
& & \partial_t^{(1)} n + \nabla_i (n u_i) = - p n \xi_n^{(1)}, \\
& & \partial_t^{(1)} u_i + \frac{k_B}{m n} \nabla_i (n T) + u_j \nabla_j u_i = - p v_T \xi_{u_i}^{(1)}, \qquad i=1,\ldots,d, \\
& & \partial_t^{(1)} T + u_i \nabla_i T + \frac{2}{d} T \nabla_i u_i = - p T \xi_T^{(1)}, \label{balanceequations1energy}
\end{eqnarray}\label{balanceequations1}
\end{subequations}
where the decay rates are given by
\begin{subequations}
\begin{eqnarray}
\xi_n^{(1)} &=& \frac{2}{n} \omega[f^{(0)},f^{(1)}], \\
\xi_{u_i}^{(1)} &=& \frac{1}{n v_T} \omega[f^{(0)},V_i f^{(1)}] + \frac{1}{n v_T} \omega[f^{(1)},V_i f^{(0)}], \qquad i=1,\ldots,d, \\
\xi_T^{(1)}&=& - \frac{2}{n} \omega[f^{(0)},f^{(1)}] + \frac{m}{n k_B T d} \omega[f^{(0)}, V^2 f^{(1)}] + \frac{m}{n k_B T d} \omega[f^{(1)},V^2 f^{(0)}].
\end{eqnarray}\label{decay1}
\end{subequations}

By definition we know that $f^{(1)}$ must be of first order in the gradients of the hydrodynamic fields; therefore for a low density gas~\cite{garzo2}
\begin{equation}
f^{(1)} = \mathcal{A}_i \nabla_i \ln T + \mathcal{B}_i \nabla_i \ln n + \mathcal{C}_{ij} \nabla_j u_i. \label{f1}
\end{equation}
The coefficients $\mathcal{A}_i$, $\mathcal{B}_i$, and $\mathcal{C}_{ij}$ depend on the fields $n$, $\mathbf{V}$, and $T$. Inserting Eq.~(\ref{f1}) in Eq.~(\ref{boltzmann1}) and making use of Eqs.~(\ref{normalsolution}), (\ref{eqscaling}), and~(\ref{balanceequations0}), one obtains the following set of equations for $\mathcal{A}_i$, $\mathcal{B}_i$, and $\mathcal{C}_{ij}$ (see Appendix~\ref{appendix1}):
\begin{subequations}
\begin{eqnarray}
\left\{ - p \left[ \xi_T^{(0)} T \partial_T + \xi_n^{(0)} n \partial_n + \tfrac{1}{2} \xi_T^{(0)} \right] + (J-p\Omega) \right\} \mathcal{A}_i - p \tfrac{1}{2} \xi_n^{(0)} \mathcal{B}_i &=& A_i, \label{integralequations1} \\
\left\{ - p \left[ \xi_T^{(0)} T \partial_T + \xi_n^{(0)} n \partial_n + \xi_T^{(0)} \right] + (J-p\Omega) \right\} \mathcal{B}_i - p \xi_T^{(0)} \mathcal{A}_i &=& B_i, \label{integralequations2} \\
\left\{ - p \left[ \xi_T^{(0)} T \partial_T + \xi_n^{(0)} n \partial_n \right] + (J-p\Omega) \right\} \mathcal{C}_{ij} &=& C_{ij}, \label{integralequations3}
\end{eqnarray}\label{integralequations}
\end{subequations}
where
\begin{subequations}
\begin{eqnarray}
A_i &=& \frac{V_i}{2} \frac{\partial}{\partial V_j} [V_j f^{(0)}] - \frac{k_B T}{m} \frac{\partial f^{(0)}}{\partial V_i}, \label{coeff1} \\
B_i &=& - V_i f^{(0)} - \frac{k_B T}{m} \frac{\partial f^{(0)}}{\partial V_i}, \label{coeff2} \\
C_{ij} &=& \frac{\partial}{\partial V_i}[V_j f^{(0)}] - \frac{1}{d} \frac{\partial}{\partial V_k}[ V_k f^{(0)} ]\delta_{ij}, \label{coeff3}
\end{eqnarray}\label{coeff}
\end{subequations}
and $\Omega$ is a linear operator defined by
\begin{equation}
\Omega g = f^{(0)} \xi_n^{(1)}[f^{(0)},g] - \frac{\partial f^{(0)}}{\partial V_i} v_T \xi_{u_i}^{(1)}[f^{0},g] + \frac{\partial f^{(0)}}{\partial T} T \xi_T^{(1)}[f^{(0)},g],\label{defOmega}
\end{equation}
where $g$ is either $\mathcal{A}_i$, $\mathcal{B}_i$, or $\mathcal{C}_{ij}$, and the functionals $\xi_n^{(1)}$, $\xi_{u_i}^{(1)}$, and $\xi_T^{(1)}$ are obtained from Eqs.~(\ref{decay1}) upon replacing $f^{(1)}$ by $g$.
It is possible to show that from Eqs~(\ref{coeff}) the solubility conditions
ensuring the existence of the functions $\mathcal{A}_i$, $\mathcal{B}_i$, and
$\mathcal{C}_{ij}$ are satisfied (see, e.g., \cite{garzo2}).

\subsection{Navier-Stokes transport coefficients}
%-----------------------------------------
The hydrodynamic description of the flow requires the knowledge of transport coefficients. The concern of the present section is to determine the form and coefficients of the constitutive equations. This can thus be achieved by linking those macroscopic transport coefficients with their microscopic definition. Using a first order Sonine polynomial expansion, it is then possible to find explicitly the transport coefficients to first order. 
This will allow us to express the functions $\mathcal{A}_i$, $\mathcal{B}_i$, and $\mathcal{C}_{ij}$ in terms of the transport coefficients, thus determining the distribution function $f^{(1)}$.

The pressure tensor may be put in the form
\begin{equation}
P_{ij}(\mathbf{r},t) = p^{(0)} \delta_{ij} - \eta \left(  \nabla_i u_j + \nabla_j u_i - \frac{2}{d} \delta_{ij} \nabla_k u_k \right) - \zeta \delta_{ij} \nabla_k u_k,\label{pressurephenomenological}
\end{equation}
where $p^{(0)}=n k_B T$ is the ideal gas pressure, and $\eta$ is the shear
viscosity. For a low density gas, the bulk viscosity $\zeta$ vanishes; therefore the last term in the pressure tensor may be neglected~\cite{garzo2,garzo3,duftycondmat}. Fourier's linear law for heat conduction is
\begin{equation}
q_i = - \kappa \nabla_i T - \mu \nabla_i n,\label{heatphenomenological}
\end{equation}
where $\kappa$ is the thermal conductivity and $\mu$ a transport coefficient
that has no analogue in the elastic case. A similar quantity appears for
granular gases, which again is non-vanishing in the inelastic case only~\cite{premier,Mareschal}.

The identification of Eq.~(\ref{pressurephenomenological}) with Eq.~(\ref{pressure}) using the result of the first order calculation yields
\begin{equation}
P_{ij}^{(1)} = \int_{\mathbb{R}^d} d \mathbf{v} \, D_{ij}(\mathbf{V}) f^{(1)}.\label{pressure1}
\end{equation}
Similarly, the identification of Eq.~(\ref{heatphenomenological}) with Eq.~(\ref{heat}) using the first order calculation leads to
\begin{equation}
q_i^{(1)} = \int_{\mathbb{R}^d} d \mathbf{v} \, S_i(\mathbf{V}) f^{(1)}. \label{heat1}
\end{equation}
The main steps of the calculation are shown in Appendix~\ref{appendix2}, and the result is
\begin{subequations}
\begin{eqnarray}
\eta^* = \frac{\eta}{\eta_0} &=& \frac{1}{\nu_\eta^* - \frac{1}{2} p \xi_T^{(0)*}}, \label{transportcoefficientseta} \\
\kappa^* = \frac{\kappa}{\kappa_0} &=& \frac{1}{\nu_\kappa^* - 2 p \xi_T^{(0)*}} \left[ \frac{1}{2} p \xi_n^{(0)*} \mu^* + \frac{d-1}{d}(2 a_2 + 1)\right], \label{transportcoefficientskappa} \\
\mu^* = \frac{n \mu}{T \kappa_0} &=& \frac{2}{2 \nu_\mu^* - 3 p \xi_T^{(0)*} - 2 p \xi_n^{(0)*}} \left[ p \xi_T^{(0)*} \kappa^* + \frac{d-1}{d} a_2 \right], \label{transportcoefficientsmu}
\end{eqnarray}\label{transportcoefficients}
\end{subequations}
where $a_2$ is the kurtosis of the distribution
\begin{equation}
a_2 = \frac{4}{d(d+2)} \frac{1}{v_T^4 n} \int_{\mathbb{R}^d} d \mathbf{V} \, f^{(0)}(V) - 1,
\end{equation}
and
\begin{equation}
\kappa_0 = \frac{d(d+2)}{2(d-1)}\frac{k_B}{m} \eta_0, \label{kappa0}
\end{equation}
\begin{equation}
\eta_0 = \frac{d+2}{8} \frac{\Gamma(d/2)}{\pi^{(d-1)/2}} \frac{\sqrt{m k_B T}}{\sigma^{d-1}}, \label{eta0}
\end{equation}
are the thermal conductivity and shear viscosity coefficients for hard spheres, respectively~\cite{ferziger}. $\xi_n^{(0)*} = \xi_n^{(0)}/\nu_0$ and $\xi_T^{(0)*} = \xi_T^{(0)}/\nu_0$ are the dimensionless decay rates, where $\nu_0 = p^{(0)}/\eta_0$, with $p^{(0)} = n k_B T$. The dimensionless coefficients $\nu_\eta^*$, $\nu_\kappa^*$, and $\nu_\mu^*$ are given by
\begin{subequations}
\begin{eqnarray}
\nu_\kappa^* &=& \frac{1}{\nu_0} \frac{\int_{\mathbb{R}^d} d \mathbf{V} \, S_{i}(\mathbf{V}) J \mathcal{A}_{i}}{\int_{\mathbb{R}^d} d \mathbf{V} \, S_{i}(\mathbf{V}) \mathcal{A}_{i}} - p \frac{1}{\nu_0} \frac{\int_{\mathbb{R}^d} d \mathbf{V} \, S_{i}(\mathbf{V}) \Omega \mathcal{A}_{i}}{\int_{\mathbb{R}^d} d \mathbf{V} \, S_{i}(\mathbf{V}) \mathcal{A}_{i}}, \\
\nu_\mu^* &=& \frac{1}{\nu_0} \frac{\int_{\mathbb{R}^d} d \mathbf{V} \, S_{i}(\mathbf{V}) J \mathcal{B}_{i}}{\int_{\mathbb{R}^d} d \mathbf{V} \, S_{i}(\mathbf{V}) \mathcal{B}_{i}} - p \frac{1}{\nu_0} \frac{\int_{\mathbb{R}^d} d \mathbf{V} \, S_{i}(\mathbf{V}) \Omega \mathcal{B}_{i}}{\int_{\mathbb{R}^d} d \mathbf{V} \, S_{i}(\mathbf{V}) \mathcal{B}_{i}}, \\
\nu_\eta^* &=& \frac{1}{\nu_0} \frac{\int_{\mathbb{R}^d} d \mathbf{V} \, D_{ij}(\mathbf{V}) J \mathcal{C}_{ij}}{\int_{\mathbb{R}^d} d \mathbf{V} \, D_{ij}(\mathbf{V}) \mathcal{C}_{ij}} - p\frac{1}{\nu_0} \frac{\int_{\mathbb{R}^d} d \mathbf{V} \, D_{ij}(\mathbf{V}) \Omega \mathcal{C}_{ij}}{\int_{\mathbb{R}^d} d \mathbf{V} \, D_{ij}(\mathbf{V}) \mathcal{C}_{ij}}.
\end{eqnarray}\label{bigintegrals}
\end{subequations}

It must be emphasized that the above results are still exact within the
Chapman-Enskog expansion framework. However, the
relations~(\ref{bigintegrals}) and the decay rates~(\ref{decay0}) cannot be
evaluated analytically without approximations. For this purpose, we first
consider the Sonine expansion for $f^{(0)}$. It was shown that to first non-Gaussian contribution in Sonine polynomials the distribution $f^{(0)}$ reads~\cite{coppex}
\begin{equation}
f^{(0)}(\mathbf{V}) = \frac{n}{v_T^d} \mathcal{M}\left( \frac{V}{v_T} \right) \left\{ 1 + a_2 \left[ \frac{1}{2} \frac{V^4}{v_T^4} - \frac{d+2}{2} \frac{V^2}{v_T^2} + \frac{d(d+2)}{8} \right]  \right\} \label{f0sonine}
\end{equation}
where
\begin{equation}
\mathcal{M}\left( \frac{V}{v_T} \right) = \frac{1}{\pi^{d/2}} \mathrm{e}^{-V^2/v_T^2} \label{maxwellian}
\end{equation}
is the Maxwellian and
\begin{equation}
a_2 = 8 \frac{3 - 2 \sqrt{2}}{4 d + 6 - \sqrt{2} + \frac{1-p}{p} 8 \sqrt{2} (d-1)}. \label{a2}
\end{equation}
The coefficient $a_2$ was shown to be in very good agreement with direct Monte Carlo simulations~\cite{coppex}. The relation~(\ref{f0sonine}) allows us to compute the decay rates (see Appendix~\ref{appendix3}):
\begin{subequations}
\begin{eqnarray}
\xi_n^{(0)*} &=& \frac{d+2}{4} \left( 1- a_2 \frac{1}{16} \right), \label{decay0nend} \\
\xi_T^{(0)*} &=& \frac{d+2}{8 d} \left( 1 + a_2 \frac{8d+11}{16} \right). \label{decay0tend}
\end{eqnarray}\label{decay0end}
\end{subequations}
Next, we retain only the first order in a Sonine polynomial expansion applied to $\boldsymbol{\mathcal{A}}$, $\boldsymbol{\mathcal{B}}$, and $\boldsymbol{\mathcal{C}}$. We thus have
\begin{subequations}
\begin{eqnarray}
\boldsymbol{\mathcal{A}}(\mathbf{V}) &=& a_1 \mathcal{M}(\mathbf{V}) \mathbf{S}(\mathbf{V}), \\ \boldsymbol{\mathcal{B}}(\mathbf{V}) &=& b_1 \mathcal{M}(\mathbf{V}) \mathbf{S}(\mathbf{V}), \\
\boldsymbol{\mathcal{C}}(\mathbf{V}) &=& c_0 \mathcal{M}(\mathbf{V}) \mathbf{D}(\mathbf{V}),
\end{eqnarray}\label{firstorderexpansion}
\end{subequations}
where $a_1$, $b_1$, and $c_0$ are the coefficients of the development. This allows us to compute the relations~(\ref{bigintegrals}). For this purpose, as already shown the probabilistic collision operator $J$ given by Eq.~(\ref{opcollision}) can be split into the sum of an annihilation operator and of a collision operator. Each contribution may thus be treated separately. Therefore we make use of previous calculations for the collision process~\cite{breycubero}. The calculations for the annihilation operator are shown in Appendix~\ref{appendix4}, and the final results read
\begin{subequations}
\begin{eqnarray}
\nu_\kappa^* = \nu_\mu^* &=& p \frac{1}{32 d} \left[ 16 + 27d + 8d^2 + a_2 \frac{2880 +1544 d - 2658 d^2 - 1539 d^3 - 200 d^4}{32 d (d+2)} \right] \nonumber \\
& & + (1-p) \frac{d-1}{d} \left( 1 + a_2 \frac{1}{32} \right), \\
\nu_\eta^* &=& p \frac{1}{8 d} \left[ 3 + 6 d + 2 d^2 - a_2 \frac{278 + 375 d + 96 d^2 + 2 d^3}{32 (d+2)} \right] + (1-p) \left( 1 - a_2 \frac{1}{32} \right).
\end{eqnarray}\label{transportend}
\end{subequations}
One may check that these expressions approach the known elastic values when $p\to 0$]\cite{breycubero}.
The transport coefficients are thus found from Eqs.~(\ref{transportcoefficients}) using Eqs.~(\ref{kappa0}), (\ref{eta0}), (\ref{a2}), (\ref{decay0end}), and~(\ref{transportend}).

In order to establish the decay rates to first order, one needs the distribution $f^{(1)}$ (see Appendix~\ref{appendix5}):
\begin{equation}
f^{(1)}(\mathbf{r},\mathbf{V};t) = - \frac{\beta^3}{n} \mathcal{M}(\mathbf{V}) \left[ \frac{2m}{d+2} S_i(\mathbf{V}) \left( \kappa \nabla_i T + \mu \nabla_i n\right) + \frac{\eta}{\beta} D_{ij}(\mathbf{V}) \nabla_j u_i \right]. \label{f1end}
\end{equation}

\subsection{Hydrodynamic equations}
%-----------------------------------
The pressure tensor and the heat flux defined by Eqs.~(\ref{pressurephenomenological}) and~(\ref{heatphenomenological}), respectively, are of order $1$ in the gradients. Thus their insertion in the balance equations~(\ref{balanceequations}) yields contributions of order $2$ in the gradients. Consequently there are second order terms (so called Burnett order) that contribute to the first order (so called Navier-Stokes order) transport coefficients, and the knowledge of the distribution $f^{(2)}$ is thus necessary. Indeed, use was made of the zeroth order relations $P_{ij} = p^{(0)} \delta_{ij}$ and $q_i = 0$ to establish the balance equation for energy~(\ref{balanceequations1energy}). However, it was shown in the framework of the weakly inelastic gas -- and consequently for an elastic gas -- that those Burnett contributions were three orders of magnitude smaller than the Navier-Stokes contributions~\cite{premier}. For the sake of simplicity, we will here neglect those second order contributions. For small annihilation probabilities $p$, this approximation is thus likely to be justified. However, we have \emph{a priori} no control on the error made when the annihilation probability $p$ is close to unity.

The hydrodynamic Navier-Stokes equations are given by
\begin{subequations}
\begin{eqnarray}
& & \partial_t n + \nabla_i (n u_i) = - p n [\xi_n^{(0)} + \xi_n^{(1)}], \\
& & \partial_t u_i + \frac{1}{m n} \nabla_j P_{ij} + u_j \nabla_j u_i = - p v_T \xi_{u_i}^{(1)}, \qquad i=1,\ldots,d, \\
& & \partial_t T + u_i \nabla_i T + \frac{2}{n k_B d}( P_{ij} \nabla_i u_j + \nabla_i q_i) = - p T [\xi_T^{(0)} + \xi_T^{(1)}],
\end{eqnarray}\label{balanceequationsend}
\end{subequations}
where the decay rates $\xi_n^{(0)}$ and $\xi_T^{(0)}$ are given by Eqs.~(\ref{decay0nend}) and~(\ref{decay0tend}), respectively. $P_{ij}$ and $q_j$ are given by Eqs.~(\ref{pressurephenomenological}) with $\zeta=0$, and~(\ref{heatphenomenological}) respectively. The rates $\xi_n^{(1)}$, $\xi_{u_i}^{(1)}$, and $\xi_T^{(1)}$ may be calculated using their definition~(\ref{balanceequations1}) and the distribution~(\ref{f1end}). We find (see Appendix~\ref{appendix6}):
\begin{subequations}
\begin{eqnarray}
\xi_n^{(1)} &=& 0 , \\
\xi_{u_i}^{(1)} &=& - v_T \left( \kappa^* \frac{1}{T} \nabla_i T + \mu^* \frac{1}{n} \nabla_i n \right) \xi_{u}^*,\\
\xi_T^{(1)}&=& 0,
\end{eqnarray}\label{decay1end}
\end{subequations}
where
\begin{equation}
\xi_u^* = \frac{(d+2)^2}{32(d-1)} \left[1 + a_2 \frac{-86 -101 d + 32 d^2 + 88d^3 + 28d^4}{32 (d+2)} \right].
\end{equation}
We thus have a closed set of equations for the hydrodynamic fields to the Navier-Stokes order.

%=========================================================
\section{Stability analysis}\label{section4}
%=========================================================
The hydrodynamic Eqs.~(\ref{balanceequationsend}) form a set of first order nonlinear partial differential equations that cannot be solved analytically in general. 
However, their linear stability analysis allows us to answer the question of formation of inhomogeneities. The scope of the present study is to find under which conditions the homogeneous solution to zeroth order, i.e., the HCS, is unstable under spatial perturbations. To this end we consider a small deviation from the HCS and the linearization of Eqs.~(\ref{balanceequationsend}) in the latter perturbation. Equations~(\ref{balanceequations0}) give the time evolution of the HCS, which is found to be
\begin{subequations}
\begin{eqnarray}
n_H(t) &=& n_0 \left(1+p \frac{t}{t_0} \right)^{-\gamma_n}, \label{stability1} \\
T_H(t) &=& T_0 \left(1+p \frac{t}{t_0} \right)^{-\gamma_T}, \label{stability2}
\end{eqnarray}
\end{subequations}
where the decay exponents are $\gamma_n = \xi_n^{(0)}(0)   t_0$, $\gamma_T =
\xi_T^{(0)}(0) t_0$, and the relaxation time $t_0^{-1} = \xi_n^{(0)}(0) +
\xi_T^{(0)}(0)/2$. The subscript $H$ denotes a quantity evaluated in the
homogeneous state. The density and temperature fields of the HCS are thus
decreasing monotonically in time, with exponents that depend on the
annihilation probability through the kurtosis of the velocity distribution. 
The explicit expression of the decay exponents may be obtained straightforwardly using Eqs.~(\ref{decay0end}).

\begin{figure}
\begin{center}
\psfrag{sp1}{$s_\perp$}
\psfrag{sp2}{$s_\parallel$}
\includegraphics[width=0.7\columnwidth]{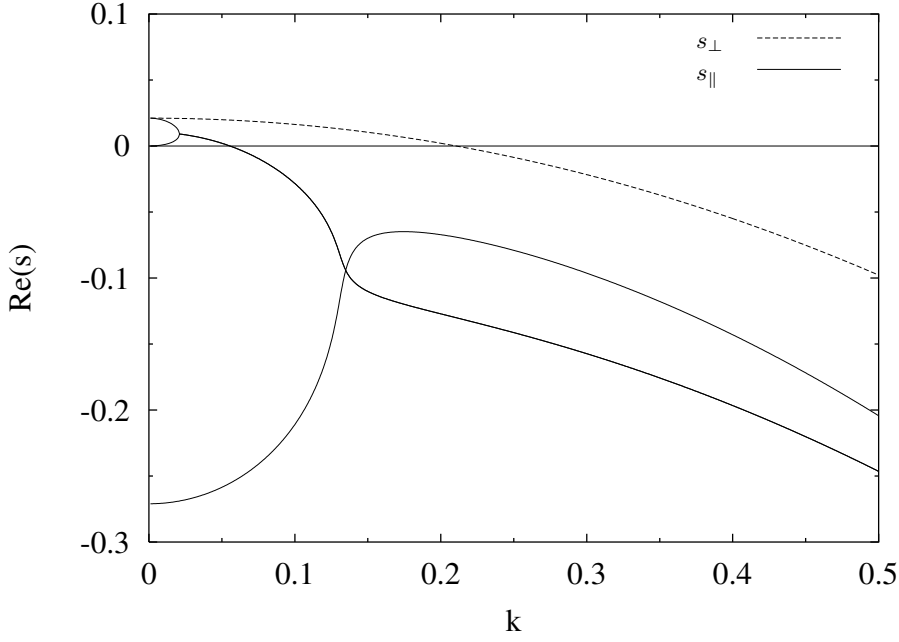}
\end{center}
\caption{Real part of the eigenvalues in dimensionless units for probabilistic ballistic annihilation with $p=0.1$ and $d=3$. The dispersion relation obtained from Eq.~(\ref{modeperp}) is represented by a dashed line (labeled $s_\perp$) whereas the three remaining relations obtained upon solving Eq.~(\ref{matrix}) are represented by continuous lines (labeled $s_\parallel$).}
\label{fig1}
\end{figure}

The linearization procedure used here follows the same route as the method used for granular gases~\cite{premier}. We define the deviations of the hydrodynamic fields from the HCS by
\begin{equation}
\delta y(\mathbf{r},t) = y(\mathbf{r},t) - y_H(t), \label{stability3}
\end{equation}
where $y=\{ n, \mathbf{u}, T\}$. Inserting the form~(\ref{stability3}) in Eqs.~(\ref{decay1end}) yields differential equations with time-dependent
coefficients. In order to obtain coefficients that do not depend on time, it is necessary to introduce the new dimensionless space and time scales defined by
\begin{subequations}
\begin{eqnarray}
\mathbf{l} &=& \frac{1}{2} \nu_{0H}(t) \sqrt{\frac{m}{k_B T_H(t)}} \mathbf{r}, \label{stability4a} \\
\tau &=& \frac{1}{2} \int_0^t d s \, \nu_{0H}(s), \label{stability4b}
\end{eqnarray}\label{stability4}
\end{subequations}
as well as the dimensionless Fourier fields
\begin{subequations}
\begin{eqnarray}
\rho_\mathbf{k}(\tau) &=& \frac{\delta n_\mathbf{k}(\tau)}{n_H(\tau)}, \label{stability6a} \\
\mathbf{w}_\mathbf{k}(\tau) &=& \sqrt{\frac{m}{k_B T_H(\tau)}} \delta \mathbf{u}_\mathbf{k}(\tau), \label{stability6b} \\
\theta_\mathbf{k}(\tau) &=& \frac{\delta T_\mathbf{k}(\tau)}{T_H(\tau)}, \label{stability6c}
\end{eqnarray} \label{stability6}
\end{subequations}
where
\begin{equation}
\delta y_\mathbf{k}(\tau) = \int_{\mathbb{R}^d} d \mathbf{l} \, \mathrm{e}^{-i \mathbf{k} \cdot \mathbf{l}} \delta y(\mathbf{l},\tau). \label{stability7}
\end{equation}
From Eq. (\ref{stability4a}), it appears that lengths are made dimensionless
making use of the time dependent  mean free path as a reference scale.
Making use of Eqs.~(\ref{stability6}) and~(\ref{stability4}) in Eqs.~(\ref{balanceequationsend}), the linearized hydrodynamic equations read
\begin{subequations}
\begin{eqnarray}
\left[ \frac{\partial}{\partial \tau} + 2 p \xi_n^{(0)*} \right] \rho_{\mathbf{k}}(\tau) + p \xi_n^{(0)*} \theta_\mathbf{k}(\tau) + i k w_{\mathbf{k}_\parallel}(\tau) =0, \label{stability8a} \\
\left[ \frac{\partial}{\partial \tau} - p \xi_T^{(0)*} + \frac{d-1}{d} \eta^* k^2 \right] \mathbf{w}_{\mathbf{k}_\parallel} + i \mathbf{k} \Big[ \big( 1-p\xi_u^* \mu^* \big) \rho_\mathbf{k}(\tau) + \big( 1-p\xi_u^* \kappa^* \big) \theta_\mathbf{k}(\tau) \Big] = 0, \label{stability8b} \\
\left[ \frac{\partial}{\partial \tau} - p \xi_T^{(0)*} + \frac{1}{2} \eta^* k^2 \right] \mathbf{w}_{\mathbf{k}_\perp}(\tau) = 0, \label{stability8c}\\
\left[ \frac{\partial}{\partial \tau} + p \xi_T^{(0)*} + \frac{d+2}{2(d-1)} \kappa^* k^2 \right] \theta_\mathbf{k}(\tau) + \left[ 2 p \xi_T^{(0)*} + \frac{d+2}{2(d-1)} \mu^* k^2 \right] \rho_\mathbf{k}(\tau) + \frac{2}{d} i k w_{\mathbf{k}_\parallel}(\tau) = 0, \label{stability8d}
\end{eqnarray} \label{stability8}
\end{subequations}
where $\mathbf{w}_{\mathbf{k}_\parallel}$ and $\mathbf{w}_{\mathbf{k}_\perp}$
are the longitudinal and transverse part of the velocity vector defined by
$\mathbf{w}_{\mathbf{k}_\parallel} = (\mathbf{w}_{\mathbf{k}} \cdot
\widehat{\mathbf{e}}_\mathbf{k})\widehat{\mathbf{e}}_\mathbf{k}$ and
$\mathbf{w}_{\mathbf{k}_\perp} = \mathbf{w}_{\mathbf{k}} -
\mathbf{w}_{\mathbf{k}_\parallel}$, where $\widehat{\mathbf{e}}_\mathbf{k}$ is
the unit vector along the direction given by
$\mathbf{k}$. Eq.~(\ref{stability8c}) for the shear mode is decoupled from the other equations and can be integrated directly so that
\begin{equation}
\mathbf{w}_{\mathbf{k}_\perp}(\tau) = \mathbf{w}_{\mathbf{k}_\perp}(0) \exp[s_\perp(p,k) \tau], \label{solperp}
\end{equation}
where
\begin{equation}
s_\perp(p,k) = p \xi_T^{(0)*} - \frac{1}{2} \eta^* k^2. \label{modeperp}
\end{equation}
The transversal velocity field $\mathbf{w}_{\mathbf{k}_\perp}$ lies in the $(d-1)$ dimensional vector space that is orthogonal to the vector space generated by $\mathbf{k}$, and therefore the mode $s_\perp$ identifies $(d-1)$ degenerated perpendicular modes. The longitudinal velocity field $\mathbf{w}_{\mathbf{k}_\parallel}$ lies in the vector space of dimension $1$ generated by $\mathbf{k}$. Hence there are three hydrodynamic fields to be determined, namely the density $\rho_\mathbf{k}$, temperature $\theta_\mathbf{k}$, and longitudinal velocity field $\mathbf{w}_{\mathbf{k}_\parallel} = w_{\mathbf{k}_\parallel} \widehat{\mathbf{e}}_\mathbf{k}$. The linear system thus reads
\begin{equation}
\begin{pmatrix}
\dot{\rho}_\mathbf{k} \\
\dot{w}_{\mathbf{k}_\parallel} \\
\dot{\theta}_\mathbf{k}
\end{pmatrix} =
\mathbf{M} \cdot \begin{pmatrix}
\rho_\mathbf{k} \\
w_{\mathbf{k}_\parallel} \\
\theta_\mathbf{k}
\end{pmatrix},
\end{equation}
with the hydrodynamic matrix
\begin{equation}
\mathbf{M} = \begin{pmatrix}
-2 p \xi_n^{(0)*} & - i k & - p \xi_{n}^{(0)*} \\
- i k (1-p \xi_u^* \mu^*) & p \xi_T^{(0)*} - \frac{d-1}{d} \eta^* k^2 & - ik(1-p \xi_u^* \kappa^*) \\
- 2 p \xi_T^{(0)*} - \frac{d+2}{2(d-1)} \mu^* k^2 & -\frac{2}{d} i k & - p \xi_T^{(0)*} - \frac{d+2}{2(d-1)} \kappa^* k^2
\end{pmatrix}. \label{matrix}
\end{equation}
The corresponding eigenmodes are given by $\varphi_n(k) = \exp[s_n(p,k) \tau]$, $n=1,\ldots,3$, where $s_n(p,k)$ are the eigenvalues of $\mathbf{M}$. Each of the three fields above is a linear combination of the eigenmodes; thus only the biggest real part of the eigenvalue $s_n(p,k)$ has to be taken into account to discuss the limit of marginal stability of the parallel mode of the velocity field. Figure.~\ref{fig1} shows the real part of the eigenvalues for $p = 0.1$ and $d=3$ (obtained numerically).

\begin{figure}
\begin{center}
\psfrag{x}{$k_\perp, k_\parallel$}
\psfrag{kp1}{$k_\perp$}
\psfrag{kp2}{$k_\parallel$}
\includegraphics[width=0.7\columnwidth]{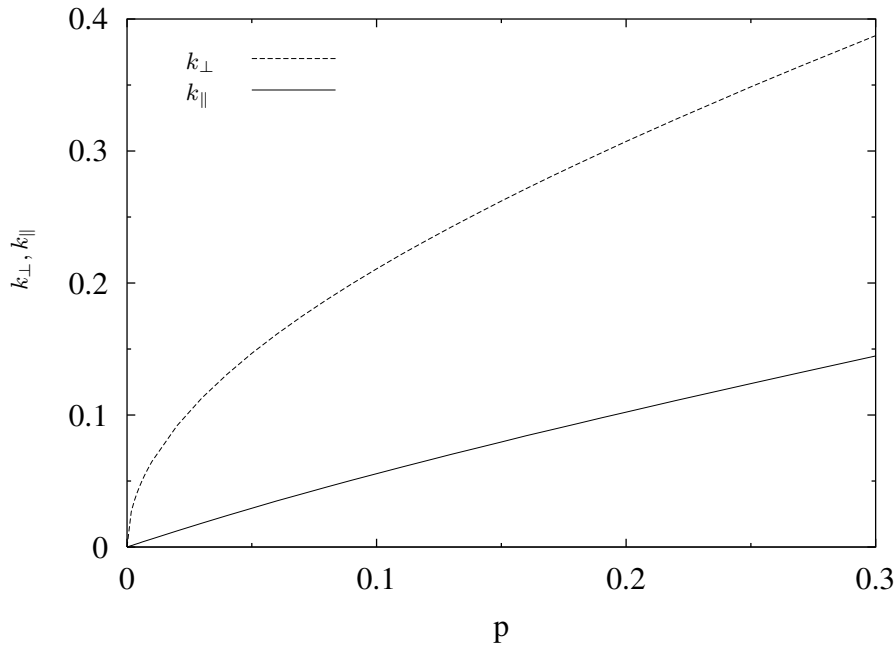}
\end{center}
\caption{Wavenumber $k_\perp$ and $k_\parallel$ in dimensionless units as a function of the annihilation probability $p$ for $d=3$.}
\label{fig2}
\end{figure}

One may identify three regions from the dispersion relations. We first define 
$k_\perp$ (dimensionless) by the condition $\mathrm{Re}[s_\perp(k_\perp,p)] = 0$, i.e.,
\begin{equation}
k_\perp = \sqrt{\frac{2 p \xi_T^{(0)*}}{\eta^*}},
\end{equation}
and $k_\parallel$ by $\max_{k_\parallel} \mathrm{Re} [s_\parallel(k_\parallel,p)] = 0$ (the expression for $k_\parallel$ is too cumbersome to be given here); we have $k_\parallel < k_\perp$. Figure~\ref{fig2} shows the dependence of $k_\perp$ and $k_\parallel$ as a function of the annihilation probability $p$. Then for all $k > k_\perp$ all eigenvalues are negative and therefore, according to Eq.~(\ref{solperp}), correspond to linearly stable modes. For $k \in [k_\parallel,k_\perp]$ only the  shear mode $\mathbf{w}_{\mathbf{k}_\perp}$ of the velocity field is linearly unstable. In the case of granular gases in dimension larger than one this region exhibits velocity vortices~\cite{goldhirschzanetti,mcnamarayoung,breymonterocubero2},
with a possible subsequent non-linear coupling to density inhomogeneities. 
From $\xi_T^{(0)*} = \xi_T^{(0)}/\nu_0$ and Eq.~(\ref{stability4b}) one may integrate Eq.~(\ref{balanceequations0c}) in order to find $T_H(\tau) = T_H(0) \exp[-2 p \xi_T^{(0)*} \tau]$. Then equating Eqs.~(\ref{stability6b}) and~(\ref{solperp}), making use of the latter expression for $T_H(\tau)$, of Eq.~(\ref{modeperp}), and of Eq.~(\ref{stability6b}) for $\tau = 0$, one finds
\begin{equation}
\delta \mathbf{u}_{\mathbf{k}_\perp}(\tau) = \mathbf{u}_{\mathbf{k}_\perp}(0) \exp \left( -\frac{1}{2} \eta^* k^2 \tau \right). \label{decaytemp}
\end{equation}
The exponential decay in the reduced variable $\tau$ translates into a power-law-like decay in the original variable $t$ [since the exponent $k=k(t)$ depends itself on time]. Indeed, the integration of Eq.~(\ref{balanceequations0c}) yields $\tau = - \ln [T_H(t)/T_H(0)] / 2 \xi_T^{(0)*}$, which we replace in Eq.~(\ref{decaytemp}) and make use of the homogeneous solution $T_H(t)$ given by Eq.~(\ref{stability2}) in order to finally obtain
\begin{equation}
\delta \mathbf{u}_{\mathbf{k}_\perp}(t) = \mathbf{u}_{\mathbf{k}_\perp}(0) \left( 1 + p \frac{t}{t_0} \right)^{-\frac{\eta^* k^2}{4 t_0^*}},
\end{equation}
where $t_0^* = t_0 / \nu_H(0)$ is the dimensionless relaxation time. In the linear approximation the perturbation of the transversal velocity field
therefore decays even if $s_\perp (k,p) > 0$. The rescaled modes with $k<k_{\parallel}$ are linearly unstable. 

However, a crucial point is that for any real (finite) system, the wave numbers
are larger than $2\pi/L$ (assuming a cubic box of size $L$), which
corresponds to a {\em time dependent} dimensionless wave number 
$k_\textrm{min} = 2\pi/(L n \sigma^{d-1})$, which 
increases with time  as $1/n$. This lower cutoff therefore inevitably enters into
the stable region $k_\textrm{min}>k_\perp$,
so that an instability may only be a transient effect. In other words,
an unstable mode associated with a given value of $k$ corresponds
to a perturbation at a wavelength which increases with time in real space,
and ultimately becomes larger than system size. However, at late times, 
the Knudsen number defined as the ratio of mean free path (which is proportional to $k_\textrm{min}$) over system size, becomes
large, which should invalidate a Navier-Stokes-like description. 
Similarly, the present coarse-grained approach is {\it a priori} restricted to low
enough values of $k$. 
Given that $k_\perp$ increases quite rapidly with $p$
(see Fig.~\ref{fig2}), the stable region $k>k_\perp$ might 
correspond to a ``non-hydrodynamic'' regime when $p$ is larger than 
some (difficult to quantify) threshold. 
Conclusions concerning the stability of the system
for such parameters rely on the validity of the hydrodynamic 
description (which could be tested by Monte Carlo or molecular
dynamics simulations) 
which is beyond the scope of the present article.

At this point, we conclude that the system may exhibit transient instabilities, but safe statements may only be made for very low values of $p$ for which $k_\perp$ is low enough to guarantee that the hydrodynamic analysis holds. The stable region is then ultimately met irrespective of system size.

\begin{figure}
\begin{center}
\psfrag{sp1}{$s_\perp$}
\psfrag{sp2}{$s_\parallel$}
\includegraphics[width=0.7\columnwidth]{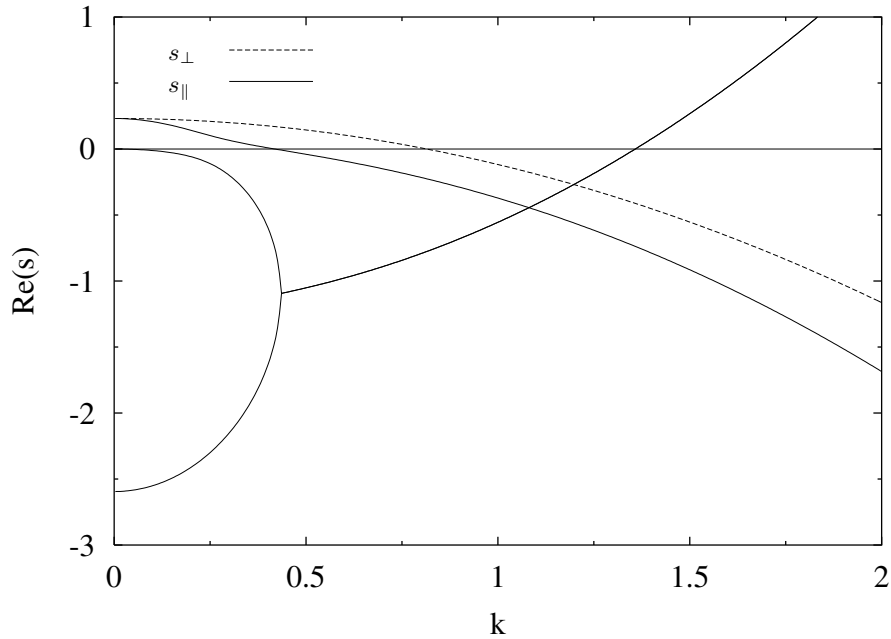}
\end{center}
\caption{Real part of the eigenvalues in dimensionless units for probabilistic ballistic annihilation with $p=0.95$ and $d=3$. The figure caption is the same as for Fig.~\ref{fig1}.}
\label{fig3}
\end{figure}

With the above possible restrictions in mind, it is instructive
to consider the counterpart of Fig.~\ref{fig1} for ``large'' 
values of $p$ (see Fig. \ref{fig3}). For $p>0.893\ldots$, we obtain the
unphysical result that some eigenvalues increase and diverge upon
increasing $k$. An a priori similar deficiency was reported for the inelastic Maxwell model where some transport coefficients ---which are obtained exactly within the Chapman-Enskog method--- diverge for strong dissipation~\cite{santosrefereeemmanuel}.

%=========================================================
\section{Conclusion}\label{section5}
%=========================================================

In this paper we construct a hydrodynamic description for probabilistic
ballistic annihilation in arbitrary dimension $d \geq 2$, where none of the
hydrodynamic fields can be associated with a conserved quantity. 
The motivation is not only to discuss the possibility
of large scale instabilities in such a system, but also
to provide the starting point for further (numerical) studies centered on the
applicability of hydrodynamics to systems in which there are 
no collisional invariants. To this aim,
we consider the low density and long-time regimes in order to make use of the
Boltzmann equation with the homogeneous cooling state (HCS) as a reference
state. The Chapman-Enskog method then allows us to build a systematic expansion
in the gradients of the fields, with an associated timescale hierarchy. We
consider only the first (Navier-Stokes) order  in the gradients to build the
hydrodynamic equations describing the dynamics of probabilistic ballistic
annihilation. The transport coefficients and decay rates are established from
the microscopic approach neglecting Burnett contributions and restricting
ourselves to the first non-Gaussian term in a Sonine expansion. We then
linearize the hydrodynamic equations around the HCS. The subsequent dispersion
relations inform on the range of the perturbation's wavelength 
and time scales for which the
system may exhibit density inhomogeneities. 

Interestingly, the behavior of the dispersion relations and of the wave numbers 
$k_\perp$ and $k_\parallel$ is qualitatively similar to its counterpart obtained
for (inelastic) granular gases~\cite{premier,breymonterocubero}. This leads us to conclude
that some features of those models do not depend on the details of the
dynamics, but rather on the parameter controlling the dissipation (referring
to the existence of non conserved quantities) in the system, namely, $p$ [or
$(1-\alpha^2)$ in the case of granular gases, where $\alpha$ is the
restitution coefficient]. However, a specific feature of our model is that 
the mean free path increases rapidly with time. Consequently,
even if the stability analysis leads us to the conclusion that this
feature drives the system in a region where the homogeneous solution
with a vanishing flow field is stable, the associated Knudsen
numbers may be too large to validate our coarse-grained approach.
At very small values of $p$, however, the stable region ($k>k_\perp$
with the notations of Sec.~\ref{section4}) should be relevant,
but then, the effects of transient instabilities in the case where the 
system is large enough to allow for $k_{\hbox{\scriptsize min}}<k_\perp$
or $k_{\hbox{\scriptsize min}}<k_\parallel$ seem difficult to assess.

Another point to emphasize is the amplitude of the dissipation in the
system, which appears through the decay rates of the hydrodynamic
fields. Again, there must be a clear separation between the macroscopic
time scales described by those decay rates, and between the microscopic
time scales. This separation of scales is required in the hydrodynamic approach
in order to make use  of the hydrodynamic fields $n$, $\mathbf{u}$, and $T$
that are associated with nonconserved quantities. The decay rates having the
dimension of the inverse of a time, their inverse thus defines a time scale. If
those decay rates increase, the associated time scales decay. In our case, we
clearly introduce three such time scales that are supposed to be macroscopic;
one for each nonconserved field. It is therefore required that the maximum of
these decay rates defines a macroscopic time scale that is much bigger than the
microscopic one. Nevertheless those decay rates increase as a function
of the annihilation probability and hence the decrease of the associated
time scale. One question that arises is to determine for which value of $p$ the
smallest time scale introduced by the decay rates is of the order of the
microscopic time scale -- which increases as a function of time because of the
decreasing density of particles remaining in the system. When this is the
case, the hydrodynamic description becomes irrelevant and one may not make use
of the fields $n$, $\mathbf{v}$, and $T$ any more. 
As the parameter $p$ controls the dissipation in the system, the question at
hand here --left for future work-- 
is reminiscent of the controversial issue of the validity 
of hydrodynamics for granular gases with ``low'' coefficients of restitution.

%=========================================================
\begin{acknowledgments} 
We acknowledge J. Piasecki, I. Bena, and J.J. Brey for useful discussions. This work was partially supported by the Swiss National Science Foundation and the French ``Centre National de la Recherche Scientifique''.
\end{acknowledgments}
%=========================================================

\appendix
%=========================================================
\section{Integral equations for $\mathcal{A}_i$, $\mathcal{B}_i$, and $\mathcal{C}_{ij}$ to first order}\label{appendix1}
%=========================================================

Using the normal solution~(\ref{normalsolution}), the scaling form~(\ref{eqscaling}), and the balance equations~(\ref{balanceequations0}) one may rewrite the right hand side of Eq.~(\ref{boltzmann1}) in such a way that
\begin{equation}
[ \partial_t^{(0)} + J ] f^{(1)} = A_i \nabla_i \ln T + B_i \nabla_i \ln n + C_{ij} \nabla_i u_j + p \Omega f^{(1)},
\end{equation}
where $V_i = v_i - u_i$,
\begin{equation}
\Omega f^{(1)} = f^{(0)} \xi_n^{(1)} - \frac{\partial f^{(0)}}{\partial V_i} v_T \xi_{u_i}^{(1)} + \frac{\partial f^{(0)}}{\partial T} T \xi_T^{(1)},\label{intermediate1}
\end{equation}
and
\begin{subequations}
\begin{eqnarray}
A_i &=& - V_i T \frac{\partial f^{(0)}}{\partial T} - \frac{k_B T}{m} \frac{\partial f^{(0)}}{\partial V_i} ,\\
B_i &=& -V_i f^{(0)} - \frac{k_B T}{m} \frac{\partial f^{(0)}}{\partial V_i},\\
C_{ij} &=& \frac{\partial}{\partial V_i}[V_j f^{(0)}]  + \frac{2}{d} T \frac{\partial f^{(0)}}{\partial T} \delta_{ij}.
\end{eqnarray}\label{intermediate1b}
\end{subequations}
The velocity dependence of $f^{(0)}$ occurs only through  $V/v_T$. Because of the normalization the temperature dependence of the function $f^{(0)}$ is of the form $T^{-d/2} \overline{f}^{(0)}(V/T^{1/2})$, and one obtains
\begin{equation}
- T \frac{\partial f^{(0)}}{\partial T} = \frac{1}{2} \frac{\partial}{\partial V_i} [V_i f^{(0)}]. \label{intermediate2}
\end{equation}
The insertion of Eq.~(\ref{intermediate2}) in Eqs.~(\ref{intermediate1b}) yields the relations~(\ref{coeff}).

Using the scaling form~(\ref{eqscaling}) and by definition of the decay rates~(\ref{decay0}) one has
\begin{equation}
\xi_n^{(0)} \sim \xi_T^{(0)} \sim n T^{1/2}.
\end{equation}
This yields the relations
\begin{equation}
T \frac{\partial \xi_{n,T}^{(0)}}{\partial T} = \frac{1}{2} \xi_{n,T}^{(0)}, \label{temp1}
\end{equation}
and
\begin{equation}
n \frac{\partial \xi_{n,T}^{(0)}}{\partial n} = \xi_{n,T}^{(0)}, \label{temp2}
\end{equation}
where $\xi_{n,T}^{(0)} = \{ \xi_n^{(0)}, \xi_T^{(0)}\}$.
Using the form~(\ref{f1}) in the left hand side of the Boltzmann equation~(\ref{boltzmann1}) and making use of the relations~(\ref{temp1}) and~(\ref{temp2}) one obtains after some calculations the relations~(\ref{integralequations}).

%=========================================================
\section{Equations for the transport coefficients}\label{appendix2}
%=========================================================
As we will apply a Sonine expansion, the symmetry properties of $\boldsymbol{\mathcal{A}}(\mathbf{V})$ and $\boldsymbol{\mathcal{B}}(\mathbf{V})$ are the same as those of $\mathbf{S}(\mathbf{V})$, whereas the properties of $\boldsymbol{\mathcal{C}}(\mathbf{V})$ are the same as those of $\mathbf{D}(\mathbf{V})$. Thus the insertion of Eq.~(\ref{f1}) in Eq.~(\ref{pressure1}) yields
\begin{equation}
P_{ij}^{(1)} = \int_{\mathbb{R}^d} d \mathbf{v} \, D_{ij}(\mathbf{V}) \mathcal{C}_{kl}(\mathbf{V}) \nabla_k u_l. \label{app2eq1}
\end{equation}
The identification of Eqs.~(\ref{app2eq1})
and~(\ref{pressurephenomenological}) 
yields~(see e.g. \cite{garzo3})
\begin{equation}
\eta = - \frac{1}{(d-1)(d+2)} \int_{\mathbb{R}^d} d \mathbf{V} \, D_{ij}(\mathbf{V}) \mathcal{C}_{ij}(\mathbf{V}).\label{app2eq2}
\end{equation}
Integrating Eq.~(\ref{integralequations3}) over $\mathbf{V}$ in $\mathbb{R}^d$ with weight $-1/[(d-1)(d+2)]  D_{ij}(\mathbf{V})$ and making use of Eq.~(\ref{app2eq2}) one obtains
\begin{equation}
\left[ - p \xi_T^{(0)} T \partial_T - p \xi_n^{(0)} n \partial_n + \nu_\eta \right] \eta = - \frac{1}{(d-1)(d+2)} \int_{\mathbb{R}^d} d \mathbf{V} \, D_{ij}(\mathbf{V}) C_{ij}. \label{app2eq3}
\end{equation}
Functional dependence analysis shows that $n \partial_n \eta = 0$ and $T \partial_T \eta = \eta/2$. Using the definitions~(\ref{definitiondij}) for $D_{ij}(\mathbf{V})$ and~(\ref{coeff3}) for $C_{ij}(\mathbf{V})$, it is possible to compute the right hand side of~(\ref{app2eq3}) which gives
\begin{equation}
\eta = \frac{1}{\nu_\eta - \frac{1}{2} p \xi_T^{(0)}} \frac{1}{d} \int_{\mathbb{R}^d} d \mathbf{V} \, m V^2 f^{(0)}. \label{app2eq4}
\end{equation}
Using the hydrostatic pressure $p^{(0)} = n k_B T$ with the definition~(\ref{deft}) for the temperature, and dividing Eq.~(\ref{app2eq4}) by $\eta_0$ we finally obtain Eq.~(\ref{transportcoefficientseta}).

The insertion of Eq.~(\ref{f1}) in Eq.~(\ref{heat1}) gives
\begin{equation}
q_i = \int_{\mathbb{R}^d} d \mathbf{V} \, S_i(\mathbf{V}) \mathcal{A}_k(\mathbf{V}) \nabla_k \ln T + \int_{\mathbb{R}^d} d \mathbf{V} \, S_i(\mathbf{V}) \mathcal{B}_k(\mathbf{V}) \nabla_k \ln n. \label{app2eq5}
\end{equation}
The identification of Eqs.~(\ref{app2eq5}) and~(\ref{heatphenomenological}) yields
\begin{subequations}
\begin{eqnarray}
\kappa &=& -\frac{1}{d T} \int_{\mathbb{R}^d} d \mathbf{V} \, S_i(\mathbf{V}) \mathcal{A}_i(\mathbf{V}), \label{app2eq6a} \\
\mu &=& -\frac{1}{d n} \int_{\mathbb{R}^d} d \mathbf{V} \, S_i(\mathbf{V}) \mathcal{B}_i(\mathbf{V}). \label{app2eq6b}
\end{eqnarray}\label{app2eq6}
\end{subequations}
The fact that $\mu \neq 0$ is due to the annihilation process. Integrating Eqs.~(\ref{coeff1}) and~(\ref{coeff2}) over $\mathbf{V}$ on $\mathbb{R}^d$ with weight $-S_i(\mathbf{V})/d$ and making use of Eq.~(\ref{app2eq6}), then making use of $T \partial_T(T \kappa) = 3 T \kappa/2$, $T \partial_T(n \mu) = 3 n \mu /2$, and $n \partial_n(T \kappa) = n \partial_n (n \mu) = 0$ obtained from functional dependence analysis, it follows
\begin{subequations}
\begin{eqnarray}
\kappa &=& \frac{1}{\nu_\kappa - 2 p \xi_T^{(0)}}\frac{1}{T} \left[ \frac{1}{2} p \xi_n^{(0)} n \mu - \frac{1}{d} \int_{\mathbb{R}^d} d \mathbf{V} \, S_i(\mathbf{V}) A_i(\mathbf{V}) \right], \\
\mu &=& \frac{1}{\nu_\mu - \frac{3}{2} p \xi_T^{(0)} - p \xi_n^{(0)}}\frac{1}{n} \left[ p \xi_T^{(0)} T \kappa - \frac{1}{d} \int_{\mathbb{R}^d} d \mathbf{V} \, S_i(\mathbf{V}) B_i(\mathbf{V}) \right].
\end{eqnarray}\label{app2eq7}
\end{subequations}
Using Eqs.~(\ref{coeff1}), (\ref{coeff2}), (\ref{definitionsi}), and~(\ref{a2}) one may calculate the integrals appearing in the right hand side of Eqs.~(\ref{app2eq7}):
\begin{subequations}
\begin{eqnarray}
\frac{1}{dT}\int_{\mathbb{R}^d} d \mathbf{V} \, S_i(\mathbf{V}) A_i(\mathbf{V}) &=& - \frac{d+2}{2} \frac{n k_B}{m \beta}(2 a_2 + 1), \\
\frac{1}{dn}\int_{\mathbb{R}^d} d \mathbf{V} \, S_i(\mathbf{V}) B_i(\mathbf{V}) &=& - \frac{d+2}{4} \frac{1}{\beta^2 m}2 a_2.
\end{eqnarray}\label{app2eq8}
\end{subequations} 
The insertion of Eqs.~(\ref{app2eq8}) in~(\ref{app2eq7}) yields Eqs.~(\ref{transportcoefficientskappa}) and~(\ref{transportcoefficientsmu}).

%=========================================================
\section{Evaluation of $\xi_n^{(0)*}$ and $\xi_T^{(0)*}$}\label{appendix3}
%=========================================================
The decay rates~(\ref{decay0n}) and~(\ref{decay0t}) may be computed using the definition~(\ref{definitionomega}) and Eqs.~(\ref{f0sonine}) and~(\ref{maxwellian}). We first change variables to $\mathbf{c}_i = \mathbf{V}_i/v_T$, $i=1,2$, then to $\mathbf{c}_{12} = \mathbf{c}_1 - \mathbf{c}_2$ and $\mathbf{C} = (\mathbf{c}_1 + \mathbf{c}_2)/2$ in order to decouple the integrals. Next, the integrals being isotropic with a symmetric weight, only even powers of the components of $\mathbf{C}$ and $\mathbf{c}_{12}$ will give nonzero contributions. Thus the terms $(\mathbf{C} \cdot \mathbf{c}_{12})^2$ in the integrals become $C^2 c_{12}^2 /d$. Finally, the resulting integrals may be computed using the following relation~\cite{trizac}: if we define
\begin{equation}
M_{np}^0 = \frac{1}{\pi^d} \int_{\mathbb{R}^{2d}} d \mathbf{c}_{12} d \mathbf{C} \, \mathrm{e}^{-c_{12}^2/2} \mathrm{e}^{-2C^2} c_{12}^n C^p,\label{app3eq1}
\end{equation}
\begin{multline}
M_{np} = \langle c_{12}^n C^p \rangle = \frac{1}{\pi^d} \int_{\mathbb{R}^{2d}} d \mathbf{c}_{12} d \mathbf{C} \, \mathrm{e}^{-c_{12}^2/2} \mathrm{e}^{-2C^2} c_{12}^n C^p \bigg[ 1 + a_2 \bigg\{ C^4 + \frac{1}{16} c_{12}^4 + \frac{d+2}{2d} C^2 c_{12}^2 \\
- (d+2) C^2 - \frac{d+2}{4} c_{12}^2 + \frac{d(d+2)}{4} \bigg\} \bigg],\label{app3eq2}
\end{multline}
then
\begin{equation}
M_{np}^0 = 2^{(n-p)/2} \frac{\Gamma[(d+n)/2] \Gamma[(d+p)/2]}{\Gamma(d/2)^2},\label{app3eq3}
\end{equation}
\begin{equation}
\frac{M_{np}}{M_{np}^0} = 1+\frac{a_2}{16 d}\left[ d(n^2+p^2) - 2 d(n+p) + 2 n p (d+2) \right].\label{app3eq4}
\end{equation}

Equations~(\ref{app3eq3}) and~(\ref{app3eq4}) may be easily verified using the relation
\begin{equation}
\int_{\mathbb{R}^d} d \mathbf{x} |\mathbf{x}|^n \mathrm{e}^{-\alpha x^2} = \frac{\pi^{d/2}}{\alpha^{(d+n)/2}} \frac{\Gamma[(d+n)/2]}{\Gamma(d/2)},\label{app3eq5}
\end{equation}
for $\alpha \in \mathbb{R}^+$. We thus obtain the decay rates to zeroth order~(\ref{decay0end}).

%=========================================================
\section{Evaluation of $\nu_\kappa^*$, $\nu_\mu^*$, and $\nu_\eta^*$}\label{appendix4}
%=========================================================
Using the first order Sonine expansion~(\ref{firstorderexpansion}), Eqs.~(\ref{bigintegrals}) reduces to
\begin{subequations}
\begin{eqnarray}
\nu_\eta^* &=& \frac{1}{\nu_0} \frac{\int_{\mathbb{R}^d} d \mathbf{V} \, D_{ij}(\mathbf{V}) J [\mathcal{M} D_{ij}]}{\int_{\mathbb{R}^2} d \mathbf{V} \, D_{ij}(\mathbf{V}) \mathcal{M}(\mathbf{V}) D_{ij}(\mathbf{V})} - p \frac{1}{\nu_0} \frac{\int_{\mathbb{R}^d} d \mathbf{V} \, D_{ij}(\mathbf{V}) \Omega [\mathcal{M} D_{ij}]}{\int_{\mathbb{R}^2} d \mathbf{V} \, D_{ij}(\mathbf{V}) \mathcal{M}(\mathbf{V}) D_{ij}(\mathbf{V})}, \\
\nu_\kappa^* = \nu_\mu^* &=& \frac{1}{\nu_0} \frac{\int_{\mathbb{R}^d} d \mathbf{V} \, S_{i}(\mathbf{V}) J [\mathcal{M} S_i]}{\int_{\mathbb{R}^2} d \mathbf{V} \, S_{i}(\mathbf{V}) \mathcal{M}(\mathbf{V}) S_i(\mathbf{V})} - p \frac{1}{\nu_0} \frac{\int_{\mathbb{R}^d} d \mathbf{V} \, S_{i}(\mathbf{V}) \Omega [\mathcal{M} S_i]}{\int_{\mathbb{R}^2} d \mathbf{V} \, S_{i}(\mathbf{V}) \mathcal{M}(\mathbf{V}) S_i(\mathbf{V})}.
\end{eqnarray}\label{app4eq1}
\end{subequations}
The denominators of Eqs.~(\ref{app4eq1}) are straightforward to compute using the formula~(\ref{app3eq5}). We thus find
\begin{subequations}
\begin{eqnarray}
\nu_\eta^* &=& \frac{\beta^2}{(d+2)(d-1) n \nu_0} \left[ \int_{\mathbb{R}^d} d \mathbf{V} \, D_{ij}(\mathbf{V}) J [\mathcal{M} D_{ij}] - p \int_{\mathbb{R}^d} d \mathbf{V} \, D_{ij}(\mathbf{V}) \Omega [\mathcal{M} D_{ij}] \right], \\
\nu_\kappa^* = \nu_\mu^* &=& \frac{2 m \beta^3}{d(d+2) n \nu_0} \left[ \int_{\mathbb{R}^d} d \mathbf{V} \, S_{i}(\mathbf{V}) J [\mathcal{M} S_i] - p \int_{\mathbb{R}^d} d \mathbf{V} \, S_{i}(\mathbf{V}) \Omega [\mathcal{M} S_i] \right].
\end{eqnarray}\label{app4eq2}
\end{subequations}
The collision operator $J$ defined by Eq.~(\ref{opcollision}) is made of the sum of an annihilation operator $L_a$ with weight $p$ and of an elastic collisional operator $L_c$ with weight $(1-p)$. Using previous calculations for $L_c$~\cite{breycubero}, we thus obtain the elastic gas contributions proportional to $(1-p)$ in the right hand side of Eqs.~(\ref{transportend}).

The following computations are technically simple, but lengthy. We shall thus only give the main steps. The annihilation contributions, written $\nu_\eta^{*a}$, $\nu_\kappa^{*a}$, and $\nu_\mu^{*a}$, are given by
\begin{subequations}
\begin{eqnarray}
\nu_\eta^{*a} &=& \frac{\beta^2}{(d+2)(d-1) n \nu_0} \int_{\mathbb{R}^d} d \mathbf{V} \, D_{ij}(\mathbf{V}) L_a [\mathcal{M} D_{ij}] + \nu_\eta^{*a'}, \\
\nu_\kappa^{*a} = \nu_\mu^{*a} &=& \frac{2 m \beta^3}{d(d+2) n \nu_0} \int_{\mathbb{R}^d} d \mathbf{V} \, S_{i}(\mathbf{V}) L_a [\mathcal{M} S_i] + \nu_\kappa^{*a'},
\end{eqnarray}\label{app4eq3}
\end{subequations}
where $L_a$ is given by Eqs.~(\ref{definitionla}) and~(\ref{definitionannihilation}), and
\begin{subequations}
\begin{eqnarray}
\nu_\eta^{*a'} &=& - \frac{\beta^2}{(d+2)(d-1) n \nu_0} \int_{\mathbb{R}^d} d \mathbf{V} \, D_{ij}(\mathbf{V}) \Omega [\mathcal{M} D_{ij}] , \\
\nu_\kappa^{*a'} = \nu_\mu^{*a'} &=& - \frac{2 m \beta^3}{d(d+2) n \nu_0} \int_{\mathbb{R}^d} d \mathbf{V} \, S_{i}(\mathbf{V}) \Omega [\mathcal{M} S_i].
\end{eqnarray}\label{app4eq3b}
\end{subequations}
Using the relation
\begin{equation}
\int_{\mathbb{R}^d} d \mathbf{v}_1 \, Y(\mathbf{v}_1) L_a[\mathcal{M} X] = \sigma^{d-1} \beta_1 \int_{\mathbb{R}^{2d}} d \mathbf{v}_1 d \mathbf{v}_2 \,  |\mathbf{v}_{12}| f^{(0)}(\mathbf{v}_1) \mathcal{M}(\mathbf{v}_2) X(\mathbf{v}_2)\left[ Y(\mathbf{v}_1) + Y(\mathbf{v}_2) \right],
\end{equation}
where $X$ and $Y$ are arbitrary functions, changing variables to $\mathbf{c}_i = \mathbf{v}_i/v_T$ for $i=1,2$, then changing variables to $\mathbf{c}_{12} = \mathbf{c}_1 - \mathbf{c}_2$ [in the following we adopt the notation $\mathbf{c}_{12} = (c_{12_1},\ldots, c_{12_d})$] and $\mathbf{C} = (\mathbf{c}_1 + \mathbf{c}_2)/2$ in order to decouple the integrals, replacing under the integral sign for symmetry reasons the relations $(\mathbf{C} \cdot \mathbf{c}_{12})^2$ by $C^2 c_{12}^2 /d$, and using
\begin{equation}
\int_{\mathbb{R}^{2d}} d \mathbf{C} d \mathbf{c}_{12} \, F(C)G(c_{12}) (\mathbf{C} \cdot \mathbf{c}_{12})^4 = \int_{\mathbb{R}^{2d}} d \mathbf{C} d \mathbf{c}_{12} \, F(C) G(c_{12}) \left( \frac{3}{d^2} C^4 c_{12}^4 - 2 d C_i^4 c_{12_j}^4 \right)
\end{equation}
where $i$ and $j$ can be chosen arbitrarily in the set $\{ 1,\ldots,d\}$ and $F$, $G$ are arbitrary isotropic integrable functions, one obtains
\begin{subequations}
\begin{eqnarray}
\nu_\eta^{*a} &=& \frac{1}{d (d-1) \pi^{d} \sqrt{2}} \frac{\Gamma(d/2)}{\Gamma\left( \frac{d+1}{2} \right)} H_1(a_2,d) + \nu_{\eta}^{*a'},\\
\nu_\kappa^{*a} = \nu_\mu^{*a} &=& \frac{1}{d \pi^{d} \sqrt{2}} \frac{\Gamma(d/2)}{\Gamma\left( \frac{d+1}{2} \right)} H_2(a_2,d) + \nu_{\kappa}^{*a'}, 
\end{eqnarray} \label{app4eq8}
\end{subequations}
where
\begin{multline}
H_k(a_2,d) = \sum_{(i,j) \in \Omega_\alpha^k} \alpha_{ij} \int_{\mathbb{R}^d} d \mathbf{C} \, \mathrm{e}^{- 2 C^2} C^i \int_{\mathbb{R}^d} d \mathbf{c}_{12} \mathrm{e}^{-c_{12}^2/2} c_{12}^{j+1} \\
+ \sum_{(i,j) \in \Omega_\gamma^k} \gamma_{ij} \int_{\mathbb{R}^d} d \mathbf{C} \, \mathrm{e}^{- 2 C^2} C^i C_1^4 \int_{\mathbb{R}^d} d \mathbf{c}_{12} \mathrm{e}^{-c_{12}^2/2} c_{12}^{j+1} c_{12_1}^4, \label{app4eq9}
\end{multline}
with $\alpha_{ij}$ and $\gamma_{ij}$ that are functions of $d$ and $a_2$,
$\Omega_\alpha^k$ and $\Omega_\gamma^k$ being the sets of allowed values for
the pairs $(i,j)$ defining the moments in the
integrals~(\ref{app4eq9}). Expressions for $\alpha_{ij}$, $\gamma_{ij}$,
$\Omega_\alpha^k$, and $\Omega_\gamma^k$ are too cumbersome to be given here. The integrals in the first sum of the right hand side of Eq.~(\ref{app4eq9}) may be computed using the formula~(\ref{app3eq5}). The integrals in the second sum may be computed using the particular case $i=j=k=l$ of the formula
\begin{equation}
M_{ijkl}^{(a)} \equiv \int_{\mathbb{R}^d} d \mathbf{x} \, |\mathbf{x}|^n \mathrm{e}^{-a x^2} x_i x_j x_k x_l = b^{(a)} \left[ \delta_{ijkl} + \frac{1}{3} (1-\delta_{ijkl})\left( \delta_{ij}\delta_{kl} + \delta_{ik} \delta_{jl} + \delta_{il} \delta_{jk} \right) \right]. \label{mic12}
\end{equation}
where
\begin{equation}
b^{(a)} = \pi^{d/2} \frac{3}{4} \frac{(d+n) (d+n+2)}{d(d+2)} \frac{\Gamma \left[ (d+n)/2 \right]}{\Gamma(d/2)} \frac{1}{a^{(d+n+4)/2}}. \label{ba}
\end{equation}
Using Eqs.~(\ref{app3eq5}) and~(\ref{ma}) we find $\nu_\eta^{*a'} = \nu_\kappa^{*a'} = 0$. The calculation can thus be performed and we obtain the first terms in the right hand side of Eqs.~(\ref{transportend}).

%=========================================================
\section{The distribution $f^{(1)}$}\label{appendix5}
%=========================================================
Using the form~(\ref{f1}) for $f^{(1)}$ and the first order Sonine expansion~(\ref{firstorderexpansion}) one has
\begin{equation}
f^{(1)}(\mathbf{r},\mathbf{V};t) = \mathcal{M}(\mathbf{V}) \left[ a_1 S_i(\mathbf{V})\nabla_i T + b_1 S_i(\mathbf{V})\nabla_i n + c_0 D_{ij}(\mathbf{V}) \nabla_j u_i \right]. \label{f1temp}
\end{equation}
The coefficients $a_1$, $b_1$, and $c_0$ may be expressed as functions of the transport coefficients, thus determining $f^{(1)}$. Eq.~(\ref{app2eq2}) in which we insert the Sonine expansion~(\ref{firstorderexpansion}) yields
\begin{eqnarray}
\eta &=& - c_0 \frac{1}{(d-1)(d+2)} \int_{\mathbb{R}^d} d \mathbf{V} \, D_{ij}(\mathbf{V}) \mathcal{M}(\mathbf{V}) D_{ij}(\mathbf{V}) \\
&=& - c_0 \frac{n}{\beta^2}, \label{coefficientappendixc0}
\end{eqnarray}
where we have made use of the definitions~(\ref{definitiondij}) and~(\ref{maxwellian}).

Proceeding in a similar way with Eq.~(\ref{app2eq6}) and~(\ref{firstorderexpansion}) it follows
\begin{eqnarray}
\kappa &=& - a_1 \frac{1}{d T} \int_{\mathbb{R}^d} d \mathbf{V} \, S_{i}(\mathbf{V}) \mathcal{M}(\mathbf{V}) S_{i}(\mathbf{V}) \\
&=& - a_1 \frac{d+2}{2} \frac{n k_B}{m \beta^2}, \label{coefficientappendixa1}
\end{eqnarray}
\begin{eqnarray}
\mu &=& - b_1 \frac{1}{d n} \int_{\mathbb{R}^d} d \mathbf{V} \, S_{i}(\mathbf{V}) \mathcal{M}(\mathbf{V}) S_{i}(\mathbf{V}) \\
&=& - b_1 \frac{d+2}{2} \frac{1}{m \beta^3}. \label{coefficientappendixb1}
\end{eqnarray}
Replacing in Eq.~(\ref{f1temp}) the coefficients $c_0$, $a_1$, and $b_1$ obtained from Eqs.~(\ref{coefficientappendixc0}), (\ref{coefficientappendixa1}), and~(\ref{coefficientappendixb1}), one obtains the distribution~(\ref{f1end}).

%=========================================================
\section{Evaluation of $\xi_n^{(1)}$, $\xi_{u_i}^{(1)}$, and $\xi_T^{(1)}$}\label{appendix6}
%=========================================================
The zeroth order and first order distributions $f^{(0)}$ and $f^{(1)}$ being known, it is possible to compute the first order decay rates~(\ref{decay1}). The procedure is similar to the calculation of the zero order decay rates of Appendix~\ref{appendix3}. We first change variables to $\mathbf{c}_i = \mathbf{V}_i/v_T$, $i=1,2$, then to $\mathbf{c}_{12} = \mathbf{c}_1 - \mathbf{c}_2$ and $\mathbf{C} = (\mathbf{c}_1 + \mathbf{c}_2)/2$ so that
\begin{equation}
\omega[A f^{(0)}, B f^{(1)}] = - n \frac{\Gamma(d/2)}{\Gamma\left( \frac{d+1}{2} \right)} \frac{\sqrt{2}}{\pi^d} \frac{d+2}{4} \left[ \frac{v_T}{2}\frac{d}{d-1} \left( \kappa^* \frac{1}{T} \nabla_i T + \mu^* \frac{1}{n} \nabla_i n \right) I_1 + \eta^* \nabla_j u_i I_2 \right],
\end{equation}
where $\omega[f,g]$ is defined by Eq.~(\ref{definitionomega}), $(A,B) = \{ (1,1), (V_2^2,1), (1,V_1^2),(V_{2_i},1), (1,V_{1_i}) \}$, $\mathbf{V}_i = (V_{i_1},\ldots, V_{i_d})$, and
\begin{equation}
I_1 = \int_{\mathbb{R}^d} d \mathbf{c}_{12} \, |\mathbf{c}_{12}| \mathrm{e}^{-c_{12}^2/2} \int_{\mathbb{R}^d} d \mathbf{C} \, \mathrm{e}^{-2 C^2} A(v_T \mathbf{c}_2) B(v_T \mathbf{c}_1) \left( c_1^2 - \frac{d+2}{2} \right) c_{1_i} \left[ 1 + a_2 S_2(c_2^2) \right], \label{int1}
\end{equation}
\begin{equation}
I_2 = \int_{\mathbb{R}^d} d \mathbf{c}_{12} \, |\mathbf{c}_{12}| \mathrm{e}^{-c_{12}^2/2} \int_{\mathbb{R}^d} d \mathbf{C} \, \mathrm{e}^{-2 C^2} A(v_T \mathbf{c}_2) B(v_T \mathbf{c}_1) \left( c_{1_i} c_{1_j} - \frac{1}{d} \delta_{ij} c_1^2 \right)  \left[ 1 + a_2 S_2(c_2^2) \right], \label{int2}
\end{equation}
In the above integrals, $\mathbf{c}_1$ and $\mathbf{c}_2$ are expressed as functions of $\mathbf{C}$ and $\mathbf{c}_{12}$. Then, in order to compute those integrals one needs the following additional relations:
\begin{equation}
M_{ij}^{(a)} \equiv \int_{\mathbb{R}^d} d \mathbf{x} \, |\mathbf{x}|^n \mathrm{e}^{-a x^2} x_i x_j = M^{(a)}  \delta_{ij},\label{mic120}
\end{equation}
where
\begin{equation}
M^{(a)} = \pi^{d/2} \frac{d+n}{2d} \frac{\Gamma \left[ (d+n)/2 \right]}{\Gamma(d/2)} \frac{1}{a^{(d+n+2)/2}}. \label{ma}
\end{equation}
Assuming summation over repeated indices, it is easy to show that
\begin{equation}
M_{jk}^{(a)} M_{ik}^{(a')} = M^{(a)} M^{(a')} \delta_{ij}, \label{sumid1}
\end{equation}
\begin{equation}
M_{kl}^{(a)} M_{kl}^{(a')} = d M^{(a)} M^{(a')}. \label{sumid2}
\end{equation}
Using the definition~(\ref{mic12}) one can show that
\begin{equation}
M_{ijkl}^{(a)} M_{kl}^{(a')} = \frac{d+2}{3} b^{(a)} M^{(a')} \delta_{ij}, \label{sumid3}
\end{equation}
\begin{equation}
M_{iklm}^{(a)} M_{jklm}^{(a')} = \frac{d+2}{3} b^{(a)} b^{(a')} \delta_{ij}, \label{sumid4}
\end{equation}
\begin{equation}
M_{klmn}^{(a)} M_{klmn}^{(a')} = \frac{d(d+2)}{3} b^{(a)} b^{(a')}. \label{sumid5}
\end{equation}
For symmetry reasons, many of the terms in the integrals~(\ref{int1})
and~(\ref{int2}) vanish upon integration. Nevertheless, the expressions are
very cumbersome and the use of symbolic computation programs is appreciable 
\cite{PB}. Equations.~(\ref{app3eq5}), (\ref{mic12}), (\ref{mic120}),
and~(\ref{sumid1})--(\ref{sumid4}) thus allow us after a lengthy but technically simple calculation to find the decay rates to first order~(\ref{decay1}).

%=========================================================


\begin{thebibliography}{}
\bibitem{bogolubov} N.N. Bogoliubov, in \emph{Studies in Statistical Mechanics}, edited by J. De Boer and G.E. Uhlenbeck, Vol. I (North-Holland Publ. Co., Amsterdam, 1962), p.5.
\bibitem{cohen} E.G.D. Cohen, in \emph{Statistical Mechanics of Equilibrium and Non-Equilibrium}, edited by J. Meixner (North-Holland Publ. Co., Amsterdam, 1962), p.140.
\bibitem{chapmancowling} S. Chapman and T.G. Cowling, \emph{The Mathematical Theory of Non-Uniform Gases} (Cambridge University Press, Cambridge, 1970).
\bibitem{santosmontaneroduftybrey} A. Santos, J.M. Montanero, J.W. Dufty, and J.J. Brey, Phys.~Rev.~E~\textbf{57}, 1644 (1998).
\bibitem{goldhirsch} I. Goldhirsch, in Granular Gases, edited by T. P\"oschel and S. Luding, \emph{Lecture Notes in Physics} {\bf 564} (Springer Verlag, Berlin, 2001).
\bibitem{Meerson} B. Meerson, T. P\"oschel, Y. Bromberg, Phys. Rev. Lett. {\bf 91} 024301, (2003).
\bibitem{BreyCub} J.J. Brey and D. Cubero, in Granular Gases, edited by T. P\"oschel and S. Luding, \emph{Lecture Notes in Physics} {\bf 564} (Springer Verlag, Berlin, 2001).
\bibitem{GarMon} V. Garz\'o and J.M. Montanero, Phys. Rev. E {\bf 68}, 041302 (2003).
\bibitem{Matthieu} T.P.C. van Noije, M.H. Ernst, E. Trizac, and I. Pagonabarraga, Phys.\ Rev.\ E {\bf 59}, 4326 (1999).
\bibitem{duftycondmat} J.W.~Dufty, Recent. Res. Devel. Stat. Phys.~\textbf{2}, 21-52 (2002) (e-print: cond-mat/0108444); J.W. Dufty, Adv. Compl. Syst.~\textbf{4}, 397-406 (2001) (e-print: cond-mat/0109215).
\bibitem{richardsonkafri} M.J.E. Richardson, J. Stat. Phys.~\textbf{89}, 777 (1997); Y. Kafri, J. Phys.~A~\textbf{33}, 2365 (2000).
\bibitem{coppex} F. Coppex, M. Droz, and E. Trizac, Phys.~Rev.~E.~\textbf{69}, 011303 (2004).
\bibitem{elskensfrisch} Y. Elskens and H.L. Frisch, Phys.~Rev.~A~\textbf{31}, 3812 (1985).
\bibitem{BenNaim} E. Ben-Naim, S. Redner, and F. Leyvraz, Phys. Rev. Lett. {\bf 70}, 1890 (1993); E. Ben-Naim, P.L.  Krapivsky, F. Leyvraz, and S. Redner, J. Phys. Chem. {\bf 98}, 7284 (1994). 
\bibitem{piasecki} J. Piasecki, Phys.~Rev.~E~\textbf{51}, 5535 (1995).
\bibitem{reydrozpiasecki} P-A. Rey, M. Droz, and J. Piasecki, Phys.~Rev.~E~\textbf{57}, 138 (1998).
\bibitem{Blythe} R.A. Blythe, M.R. Evans, and Y. Kafri, Phys. Rev. Lett. {\bf 85}, 3750 (2000).
\bibitem{Krapivsky} P.L. Krapivsky and C. Sire, Phys. Rev. Lett. {\bf 86}, 2494 (2001).
\bibitem{Lettre} E. Trizac, Phys. Rev. Lett. {\bf 88}, 160601 (2002).
\bibitem{trizac} J. Piasecki, E. Trizac, and M. Droz, Phys. Rev. E~\textbf{66}, 066111 (2002).
\bibitem{physica} F. Coppex, M. Droz, J. Piasecki, E. Trizac, and P. Wittwer, Phys.~Rev.~E~\textbf{67}, 021103 (2003).
\bibitem{breynew} J.J.~Brey and M.J.~Ruiz-Montero, Phys.~Rev.~E~\textbf{69}, 011305 (2004).
\bibitem{lanford} O.E. Lanford III, Physica~A~\textbf{106}, 70 (1981).
\bibitem{resibois} P. R\'esibois and M. de Leener, \emph{Classical Theory of Fluids} (Wiley, NY, 1977).
\bibitem{breymonterocubero2} J.J. Brey, M.J. Ruiz-Montero, and D. Cubero, Phys.~Rev.~E \textbf{54}, 3664 (1996).
\bibitem{premier} J.J. Brey, J.W. Dufty, C.S. Kim, and A. Santos, Phys. Rev. E~\textbf{58}, 4638 (1998).
\bibitem{grad} H. Grad, ``Principles of the kinetic theory of gases'', in \textit{Encyclopedia of Physics}, edited by S. Flugge (Springer, NY, 1958).
\bibitem{ggnoije} T.P.C. van Noije and M.H. Ernst, in Granular Gases, edited by T. P\"oschel and S. Luding, \emph{Lecture Notes in Physics} (Springer Verlag, Berlin, 2001), p.3.
\bibitem{ramirez} R. Ram\'irez, D. Risso, R. Soto, and P. Cordero, Phys.~Rev.~E~\textbf{62}, 2521 (2000).
\bibitem{tangoldhirsch} M.-L. Tan and I. Goldhirsch, Phys.~Rev.~Lett.~\textbf{81}, 3022 (1998).
\bibitem{breyruizmonterocubero} J.J. Brey, M.J. Ruiz-Montero, and D. Cubero, Europhys.~Lett.~\textbf{48} (4), 359--364 (1999).
\bibitem{santosrefereeemmanuel} A. Santos, Physica~A \textbf{321}, 442--466 (2003).
\bibitem{aggregation} G.F. Carnevale, Y. Pomeau, and W.R. Young, Phys.~Rev.~Lett.~\textbf{64}, 2913 (1990).
\bibitem{Trizac_coal} E. Trizac and J.-P. Hansen, Phys. Rev. Lett. {\bf 74}, 4114 (1995).
\bibitem{garzo2} V. Garz\'o and J. W. Dufty, Phys.~Rev.~E~\textbf{59}, 5895 (1999).
\bibitem{garzo3} V. Garz\'o and J.M. Montanero, Physica A \textbf{313}, 336-356 (2002).
\bibitem{Mareschal} R. Soto, M. Mareschal, and D. Risso, Phys. Rev. Lett. {\bf 83}, 5003 (1999).
\bibitem{ferziger} J. Ferziger and H. Kaper, \emph{Mathematical Theory of Transport Process in Gases}, North-Holland, Amsterdam (1972).
\bibitem{breycubero} J.J. Brey and D. Cubero, in Granular Gases, edited by T. P\"oschel and S. Luding, \emph{Lecture Notes in Physics} (Springer Verlag, Berlin, 2001), p.59.
\bibitem{goldhirschzanetti} I. Goldhirsch and G. Zanetti, Phys. Rev. Lett.~\textbf{70}, 1619 (1993).
\bibitem{mcnamarayoung} S. McNamara and W.R. Young, Phys.~Rev.~E~\textbf{50}, R28 (1994);Phys.~Rev.~E~\textbf{53}, 5089 (1996).
\bibitem{breymonterocubero} J.J. Brey, M.J. Ruiz-Montero, and D. Cubero, Phys.~Rev.~E~\textbf{60}, 3150 (1999).
\bibitem{PB} T. P\"oschel and N. V. Brilliantov, in Granular Gas Dynamics, volume 624 of Lecture Notes in Physics, pages 131-162, New York, 2003. Springer.

\end{thebibliography}
\end{document}